\documentclass[aps,twocolumn]{revtex4}
\usepackage[utf8]{inputenc}
 \usepackage{amsmath,amssymb}
\usepackage{color}
\usepackage{graphicx}
\usepackage{subfigure}
\usepackage{lipsum}
\usepackage{booktabs}
\usepackage{siunitx}
\usepackage{array}
\usepackage{epsf}
\usepackage{makecell}
\usepackage{epstopdf}
\usepackage{xcolor}
\begin{document}
\title{Resolving Galactic and Cluster Dynamics Without Dark Matter: Tsallis Entropy as the Unique Foundation of Emergent Gravity}
\author{S. Shamari\footnote{email:shahryar@hafez.shirazu.ac.ir} and
A. Sheykhi \footnote{email:asheykhi@shirazu.ac.ir}}

\address{Department of Physics, College of
Science, Shiraz University, Shiraz 71454, Iran\\
Biruni Observatory, College of Science, Shiraz University, Shiraz
71454, Iran}

\begin{abstract}
While modified entropy models-such as Barrow, Tsallis, Kaniadakis,
Power-law, Logarithmic, and R\'{e}nyi entropies-have been widely
explored in cosmological contexts, their implications on galactic
scales remain largely untested. These generalizations of the
Bekenstein-Hawking entropy encode quantum gravitational,
nonextensive, or fractal spacetime effects and can alter the
gravitational entropy-area relation. In this paper, we demonstrate
that the entropic force framework, when applied to galactic
rotation curves and the baryonic mass of galaxy clusters, uniquely
selects Tsallis entropy as the specific generalized entropy
formulation. We then extend this Tsallis modified gravity to
globular clusters to complete the structural hierarchy from
galaxies to galaxy clusters to globular clusters and to
investigate its behavior as a function of system scale. We will
show that the nonextensive parameter exhibits no correlation with
any of the macroscopic quantities characterizing gravitational
systems, such as mass, radius, temperature, or density.
Furthermore, it has previously been shown that entropy is not
well-defined within the standard thermodynamic approach to
gravity. The adoption of nonextensive statistics provides a
foundation for entanglement, thereby enabling a consistent
definition of entanglement entropy. We predict the existence of
galaxy clusters with $\delta = 1$ (i.e., clusters whose dynamics
require no dark matter), analogous to $\delta \simeq 1$ systems
already observed at globular cluster scales. At galactic scales,
by contrast, the data require $\delta > 1/2$ with values close to
$0.5$. This prediction provides a unique observational test to
discriminate Tsallis gravity from $\Lambda$CDM and MOND.
Therefore, for the entropic gravity paradigm to be consistent with
observational data across all scales-from globular clusters to
galaxies to galaxy clusters-it is inevitably required to be built
upon Tsallis entropy. This brings the entropic gravity program to
its natural conclusion: gravity, at its core, is consistently
described by nonextensive statistical mechanics.

\end{abstract}
 \maketitle
 \newpage
\section{Introduction\label{Intro}}
The quest to understand the fundamental nature of gravity, space,
and time has led to profound theoretical innovations, among which
the thermodynamic-gravity conjecture stands as a pivotal insight.
Originally formulated by Jacobson and later enriched by
Padmanabhan's emergence paradigm, this conjecture posits that
gravitational field equations-including those of General
Relativity-can be derived from thermodynamic principles applied to
spacetime horizons. In this framework, the Bekenstein-Hawking
entropy $S=A/(4G)$ plays a central role, linking the geometry of
horizons to the statistical mechanics of spacetime microstates. In
recent years, however, various quantum-gravitational,
nonextensive, and fractal-spacetime considerations have motivated
generalizations of this entropy formula, giving rise to modified
entropy prescriptions such as those of Barrow, Tsallis,
Kaniadakis, Power-law, Logarithmic and R\'{e}nyi. These extended
entropies introduce new parameters that encode departures from
standard thermodynamics and may reflect deep-seated quantum or
geometric properties of spacetime.

To date, research on modified entropies has been predominantly
cosmology-centric. Studies have explored how such corrections
alter the Friedmann equations, influence dark energy models,
modify inflationary scenarios, and leave imprints on the cosmic
microwave background. While these investigations have placed
valuable constraints on entropy parameters using large-scale
cosmological data, a critical and largely uncharted frontier
remains: how do modified entropies manifest on galactic and
sub-cosmological scales? Galaxies-and the dark matter halos that
host them-represent gravitational systems with well-defined
effective horizons and rich dynamical observables (rotation
curves, velocity dispersion profiles, baryonic mass-velocity
relations). Yet, the implications of horizon thermodynamics for
such systems have scarcely been explored, leaving open the
question of whether galactic kinematics can serve as a new,
independent test bed for quantum-gravitational entropy
corrections.

Our aim here is to bridge this gap by developing and applying a
framework to test modified entropies through galactic-scale
observations. We posit that if horizon thermodynamics underpins
gravitational dynamics, then the entropy associated with the
boundary of a galaxy or a dark matter halo-be it the Hubble radius
of a galaxy group or the radius enclosing a fixed density
contrast-should govern its equilibrium properties. using the
entropic force scenario proposed by Verlinde \cite{Ver} we derive
modified force laws and mass-velocity relations that depend
explicitly on the chosen entropy form. These relations can then be
confronted with high-precision galactic data from surveys such as
SPARC (for rotation curves), MaNGA (for stellar kinematics), and
other spatially-resolved kinematic datasets.

Although the origins of the dark matter problem trace back to the
early 1930s with the work of Zwicky and Oort, it only became a hot
research topic when Vera Rubin published her observations of
galactic rotation curves. Today, after nearly a century of diverse
astrophysical observations from the dynamics of galaxy clusters to
gravitational lensing studies and baryon acoustic oscillations a
consistent picture emerges: approximately $\%85$ of the matter in
the universe is non-baryonic. Despite extensive searches, direct
detection of dark matter particles remains elusive, motivating
serious consideration of alternative ideas
\cite{Zwicky1933,Oort1932,Rubin1970,Rubin1978,Rubin1980,Persic1996}.

Many different theories have been proposed to explain this puzzle
without non-baryonic dark matter such as Modified Gravity (MOG),
Modified Newtonian Dynamics (MOND), Carmelian theory, Cooperstock
model, etc
\cite{Moffat2006,Milgrom1983,Cooperstock2007,Carmeli1998,Carmeli2000}.

A parallel line of inquiry emerged from black hole thermodynamics,
following the discovery that black holes possess entropy
proportional to their horizon area and temperature. Jacobson
demonstrated that the Einstein field equations are nothing but an
equation of state for spacetime. Verlinde's entropic gravity
framework realizes the idea that gravity itself may be an emergent
phenomenon, with spacetime possessing intrinsic thermodynamic
properties. In this thermodynamic paradigm, the choice of entropy
functional becomes crucial. While Bekenstein-Hawking entropy leads
to standard general relativity, any alternative entropy yields
modified gravitational dynamics
\cite{Bardeen1973,Bekenstein1973,Hawking1975,Jacobson1995}.

In this work, we classify all possible entropy modifications into
two types of generalized entropies. We first examine their
performance in reproducing galactic rotation curves; in this
assessment, only Tsallis entropy succeeds. Next, we investigate
their performance in galaxy clusters. We reconstruct a well-known
model and again evaluate each modified entropy. Once more, Tsallis
entropy emerges successful. We then extend this Tsallis modified
gravity to globular clusters to complete the structural hierarchy.
After analyzing and plotting the corresponding figures, we turn to
the origin of Tsallis entropy and its theoretical implications.

This paper is structured as follows. Section II critically
examines type-II entropies, demonstrating their failure at
galactic and cluster scales. In Section III, we derive the
Tsallis-modified force law and apply it to galaxy clusters,
presenting our observational analysis of 40 clusters. We then
extend this framework to globular clusters, analyzing the velocity
dispersion profiles of 33 such systems. In Section IV, we discuss
the theoretical foundations of nonextensive statistics in
gravitational systems, drawing on the work of Chavanis and others,
and interpret the physical meaning of the Tsallis parameter
$\delta$ in terms of dynamical relaxation and hidden constraints.
Section V presents our central prediction: the existence of
dynamically relaxed galaxy clusters that are observationally dark
matter-free, offering a decisive test to distinguish Tsallis
gravity from $\Lambda$CDM and MOND. Finally, Section VI is devoted
to closing remarks. Throughout this paper we set $\hbar = c = k_B
= 1$.
\section{Modified Entropies and Newton's law of Gravity\label{Gen}}
As we mentioned, within the framework of cosmology and
gravitational theory, the investigation of different entropy
formulations based on the Verlinde paradigm constitutes a highly
active area of research and has yielded significant and intriguing
results. Within this paradigm, each entropy formulation gives rise
to a distinct emergent gravitational dynamics. All of these
entropy measures can be regarded as generalizations of the
Bekenstein-Hawking entropy, from which the standard equations of
general relativity as well as Newtonian gravity are recovered in
appropriate limits
\cite{Padmanabhan2002,Padmanabhan2005,Eling2006,Akbar2006a,Paranjape2006,Padmanabhan2007,Akbar2007,Padmanabhan2010,Wang2001}.
All of these entropy formulations can be summarized into two
distinct categories. The first class such as generalized
mass-to-horizon entropy \cite{Goh1,Goh2}, Barrow \cite{Barrow2020}
and Tsallis \cite{Tsallis1988} entropies, etc can be written as
\begin{equation} \label{S1}
\text{(i)} \quad S = \alpha R^{n-1} S_{BH} = \frac{\alpha \pi}{G}
R^{n-1} R^2 = \frac{\alpha \pi}{G} R^{n+1},
\end{equation}
where $R$ is the radius of the system. The second class such as
{Kaniadakis \cite{Kaniadakis2002,Kaniadakis2005}, R\'{e}nyi
\cite{Renyi1961} and Logarithmic entropies \cite{Das2002}}, etc
can be written as area low plus some correction terms,
\begin{equation} \label{S2}
\text{(ii)} \quad S = S_{BH} + \alpha f(A) = \frac{\pi R^2}{G} +
\alpha f(R),
\end{equation}
where $\alpha$ is a constant and $f(A)$ stands for the correction
terms. We first focus on the second type of entropies. The
absolute value of the force emerging from these entropic
contributions can be derived from the relation
\cite{Sheykhi2010,Sheykhi2011}
\begin{equation} \label{Fgen}
F = \frac{G M m}{R^2} \times 4 G\times \left. \frac{dS}{dA}
\right|_{A = 4\pi R^2} .
\end{equation}
For the (dual) Kaniadakis entropy, the modified entropy expression
up to the first-order correction term can be written as \cite{Kaniadakis2002,Kaniadakis2005}
\begin{equation} \label{SK}
S_{K} = S_{BH} \pm \frac{K^{2}}{6} S_{BH}^{3} +
\mathcal{O}(K^{4}),
\end{equation}
where $K$ is the Kaniadakis parameter. Here the plus and minus
signs correspond to the Kaniadakis and dual Kaniadakis entropies,
respectively. This modification leads to the following form of the
gravitational force,
\begin{equation} \label{FK}
F = - \frac{G M m}{R^{2}} \left( 1 \pm \gamma R^{4} \right).
\end{equation}
In the context of black hole physics, the R\'enyi entropy
associated with the black hole horizon is given by \cite{Renyi1961}
\begin{equation} \label{SRenyi}
S_{h} = \frac{1}{\lambda} \ln \left( 1 + \lambda S_{BH} \right) ,
\end{equation}
where $\lambda$ is called the R\'enyi parameter. In the limit
$\lambda \to 0$, one recovers the standard area law of black hole
entropy. This leads to the following modified form of Newton's law
of gravitation
\begin{equation} \label{FRenyi}
F = \frac{G M m}{R^2} \left( 1 + \frac{\lambda \pi R^2}{G} \right)^{-1} .
\end{equation}
This expression represents the modification of the Newtonian
gravitational force inspired by the R\'enyi entropy. Logarithmic
corrections to black hole entropy arise in the context of loop
quantum gravity due to thermal equilibrium fluctuations and
quantum fluctuations \cite{Das2002,Sheykhi2010,Mann1997,Kaul2000}.
The corrected entropy can be written as \cite{Sheykhi2010}
\begin{equation} \label{S_log}
S = \frac{A}{4 L_p^2} - \xi \ln \frac{A}{4 L_p^2} + \eta \frac{
L_p^2}{A},
\end{equation}
where $\xi$ and $\eta$ are dimensionless constants of order unity,
and $L_p$ denotes the Planck length. The corresponding modified
Newton's law is \cite{Sheykhi2010}
\begin{equation} \label{Flog}
F = \frac{G M m}{R^2} \left( 1 - \frac{\xi}{\pi}
\frac{L_p^2}{R^2}-\frac{\eta}{4 \pi^2} \frac{L_p^4}{R^4} \right) .
\end{equation}
Consequently, each of these entropic forces leads to a distinct
modification of the gravitational law. In the history of dark
matter studies, the problem of galactic rotation curves occupies a
special place and is regarded as the most direct evidence for the
existence of dark matter. The minimal expectation from any
modified theory of gravity is to successfully address this issue.
It is evident that none of the aforementioned forces is capable of
producing a flat rotation curve at large distances. For instance,
equating Eq.~(\ref{FRenyi}) to $ma$ and substituting the
rotational acceleration $a = v^2/R$, the rotational velocity of
stars derived from the R\'enyi-induced force takes the following
form
\begin{equation} \label{vRenyi}
v = \sqrt{\frac{G M}{R +{\lambda \pi R^{3}}/{G}}},
\end{equation}
which clearly fails to yield a flat galactic rotation curve. This
naturally raises the question of why a modified theory of gravity
that fails even the most elementary expectation has nevertheless
been extensively discussed in the scientific literatures. There
are two clear reasons for this. First, the analyses presented in
individual studies have typically focused on specific aspects of
the problem, without addressing all relevant facts simultaneously.
Second, a major conceptual mistake has been made, which we shall
discuss in detail in what follows. One of the most well-known
theories that has been extensively discussed in the scientific
literature is the Modified Newtonian Dynamics (MOND). One of the
earliest successes of MOND was its ability to reproduce the
empirical Tully-Fisher relation \cite{Tully1977} . Subsequently,
Stacy McGaugh demonstrated that a characteristic acceleration
discrepancy exists in gravitational systems, which constitutes the
central idea of the MOND paradigm \cite{McGaugh2016}.

Identifying a fundamental origin for the MOND theory has since
become an active area of research. A contemporary approach aims to
provide an underlying theoretical foundation for MOND while
simultaneously ensuring that the remarkable and unambiguous
success of the theory at galactic scales is preserved in the
resulting framework. Such an approach has also been pursued within
the context considered in this work. In particular, attempts have
been made to derive MOND from an entropic force perspective
\cite{Abreu2025}. However, a major mistake has been made in this line of
reasoning.

Let us consider the R\'enyi-induced force given by Eq.~(\ref{FRenyi}). By
defining
\begin{equation} \label{a0def}
a_0 = \lambda \pi M , \qquad a = \frac{G M}{R^2},
\end{equation}
the force can be written in the MOND-like form
\begin{equation} \label{Fmondlike}
F = m\, \mu\!\left( \frac{a}{a_0} \right) a ,
\end{equation}
with the interpolation function given by
\begin{equation} \label{mu}
\mu\!\left( \frac{a}{a_0} \right) = \frac{1}{1 + \frac{a_0}{a}} .
\end{equation}
If we define, as usual,
\begin{equation} \label{xdef}
x = \frac{a}{a_0} ,
\end{equation}
the interpolation function can equivalently be written as
\begin{equation}\label{mux}
\mu(x) = \frac{x}{1 + x} .
\end{equation}
A similar approach has also been adopted for other entropy
formalisms. However, a fundamental problem arises, which can only
be identified by examining the MOND theory more closely. In the
MOND framework, the force is written as
\begin{equation} \label{Fmond}
F = m\, \mu\!\left( \frac{a}{a_0} \right) a ,
\end{equation}
with the defining relation
\begin{equation} \label{monddef}
a\, \mu\!\left( \frac{a}{a_0} \right) = a_{\text{Newton}} ,
\end{equation}
where
\begin{equation} \label{anewton}
a_{\text{Newton}} \equiv \frac{G M}{R^2} .
\end{equation}
On the other hand, the actual acceleration in the MOND regime is
given by
\begin{equation} \label{amond}
a = \frac{G M_{\text{MOND}}}{R^2} .
\end{equation}
Therefore, in contrast to the previous calculations, the source
masses responsible for the acceleration are not identical in the
Newtonian and MOND descriptions. We now proceed one step further
in order to derive the corresponding mass relation.

We consider the interpolation function
\begin{equation} \label{mu2}
\mu(x) = \frac{x}{\sqrt{1 + x^2}} .
\end{equation}
The MOND relation is then given by
\begin{equation} \label{mondrel}
\frac{G M_{\text{N}}}{R^2} = \frac{G M_{\text{MOND}}}{R^2}\, \mu\!\left( \frac{G M_{\text{MOND}}}{a_0 R^2} \right) ,
\end{equation}
where $M_{\text{N}}$ is the Newtonian source mass, and
$M_{\text{MOND}}$ is the corresponding mass in the MOND framework.
Solving for $M_{\text{MOND}}$, one obtains
\begin{equation} \label{mmond}
M_{\text{MOND}} = M_{\text{N}} \sqrt{ 1 + \left( \frac{a_0 R^2}{G M_{\text{MOND}}} \right)^2 } .
\end{equation}
which is a nonlinear equation. It is clear that such a linearizing
approach within the entropic force framework cannot reproduce this
inherently nonlinear relation. (While other entropy formalisms
fail to provide an origin for MOND theory, only Tsallis entropy as
we shall see in Section \ref{MOND}  has the potential to overcome these
shortcomings).

We have seen that none of the type-II entropies discussed in the
scientific literatures are capable to yield  satisfactory results.
However, one step remains before reaching a conclusion. The
question then arises: is it possible to find a type-II entropy
that can successfully address the problem of galactic rotation
curves? Equation (\ref{Fgen}) can also be written in the following
form \cite{Sheykhi2010,Sheykhi2011}
\begin{equation} \label{Fgen2}
F = \frac{G M m}{R^2} \left[ 1 + 4 L_p^2 \frac{d \mathcal {S}}{d
A} \right] ,
\end{equation}
where $L_p^2=G$ ($\hbar=c=k_B=1$) and  $\mathcal {S}$ stands for
the correction term alone, namely
\begin{equation} \label{Scal}
\mathcal {S} \equiv \alpha f(R).
\end{equation}
The derivative of the correction term with respect to the area can
be written as
\begin{align} \label{dSdA}
\frac{d\mathcal {S}}{dA} &= \alpha \frac{\partial f(R)}{\partial
A} = \alpha \frac{\partial f(R)}{\partial R} \frac{\partial
R}{\partial A} = \frac{\alpha}{8 \pi R} \frac{\partial
f(R)}{\partial R},
\end{align}
where we used the relation between the horizon area and radius, $A = 4 \pi R^2$.
The acceleration can be written as
\begin{equation} \label{acc}
a = \frac{G M}{R^2} \left[ 1 + \frac{\alpha L_p^2}{2 \pi R}
\frac{\partial f}{\partial R} \right] ,
\end{equation}
which leads to the rotational velocity
\begin{equation} \label{v2}
v^2 = \frac{G M }{R} \left[ 1 + \frac{\alpha L_p^2}{2 \pi R}
\frac{\partial f}{\partial R} \right] .
\end{equation}
 Now, if we require that this relation yields
\begin{equation} \label{TF}
v^4 = G M a_0 ,
\end{equation}
corresponding to a flat rotation curve and in agreement with the
MOND prediction, and define
\begin{equation}
 a_0=  \frac{\alpha L_p^2}{\pi},
\end{equation}
then, the rotational velocity (\ref{v2}) can be written as
\begin{equation} \label{v22}
v^2 = \frac{G M}{R} + \frac{a_0}{2}  \frac{GM}{R^2} \frac{\partial
f}{\partial R} = \frac{G M}{R} + \frac{a_0}{2}  R ,
\end{equation}
provided we take
\begin{equation} \label{df}
  \frac{\partial f}{\partial R} = \frac{R^3}{GM} \Rightarrow  f(R) \propto
  R^4.
\end{equation}
Therefore, the correction term in the entropy is given by
$\mathcal{S}\sim R^4$. Using Eq. (\ref{v22}), we reach
\begin{equation} \label{v4}
v^4 = \left( \frac{G M}{R} \right)^2 + G M a_0 + \frac{1}{4} a_0^2 R^2 .
\end{equation}
Here, the second term reproduces the Tully-Fisher relation, while
the third term is negligible on galactic scales (note that
$a_0\sim 10^{-10}m/s$). This implies that the correction term in
the entropy must generally have a power-law dependence ($\mathcal
{S} \sim A^2 \sim R^4$) in order to reproduce the Tully-Fisher
relation and explain the flat rotation curve.

An example form of such a power-law correction can be written as
\cite{Das2008}
\begin{equation} \label{Spl}
S = k_B \frac{A}{4 L_p^2} \left[ 1 - K_\zeta A^{1-\zeta/2}
\right],
\end{equation}
where $\zeta$ is a dimensionless constant whose value is still
under debate, $k_B$ is the Boltzmann constant, and
\begin{equation} \label{Kzeta}
K_\zeta = \frac{\zeta (4 \pi)^{\zeta/2 - 1}}{(4-\zeta)
r_c^{2-\zeta}} ,
\end{equation}
with $r_c$ denoting the crossover scale. The second term in
Eq.~(\ref{Spl}) represents a power-law correction to the usual
area law of black hole entropy. This correction naturally arises
when the quantum entanglement between fields inside and outside
the horizon is taken into account, with the field wave-function
considered as a superposition of the ground and excited states
\cite{Das2008,Radicella2010}. However, there is a problem:  the
correction term reproduce $\mathcal {S} \sim A^2 \sim R^4$ when
$\zeta = 0$. In this case we have $K_\zeta =0 $, which means the
correction term vanishes. Similarly, for (dual) Kaniadakis entropy
given in Eq. (\ref{SK}), the correction term is $\mathcal {S} \sim
A^3 \sim R^6$ which cannot reproduce the Tully-Fisher relation.
This indicates a serious limitation of type-II entropies: they are
inherently incapable of reproducing a successful description of
galactic gravity and cannot account for flat rotation curves in a
physically consistent manner.

\subsection{Type-I Entropies}
We now turn to the analysis of type-I entropies. It should be
emphasized that the relation derived in the previous section for
type II entropies cannot be employed here. Indeed, for a general
entropy of the type II, we can write
\begin{equation} \label{SII}
S=\frac{A}{4 L_p^2} + \mathcal {S}(A),
\end{equation}
and the variation of the entropy is given by
\begin{equation} \label{DeltaS}
\Delta S = \frac{\partial S}{\partial A}\,\Delta A = \left(
\frac{1}{4 L_p^2} + \frac{\partial \mathcal {S}(A)}{\partial A}
\right)\Delta A .
\end{equation}
Within the entropic gravity framework, the energy associated with
the holographic screen is identified with the relativistic rest
energy of the source mass,
\begin{equation} \label{E}
E = M  ,
\end{equation}
while the number of microscopic degrees of freedom on the screen is assumed to scale with its area,
\begin{equation} \label{Q}
A = Q N .
\end{equation}
Assuming the validity of the equipartition theorem,
\begin{equation}\label{EQP}
E = \frac{1}{2} N  T ,
\end{equation}
and adopting the thermodynamic definition of the entropic force
\cite{Ver}
\begin{equation} \label{Fent}
F = T \frac{\Delta S}{\Delta x},
\end{equation}
with $\frac{\Delta S}{\Delta x}=2\pi m $ one obtains the entropic-corrected Newton's law of gravity as
\cite{Sheykhi2010}
\begin{equation}\label{Fcorr}
F = \frac{G M m}{R^2} \left[ 1 + 4 L_p^2 \frac{\partial \mathcal
{S}}{\partial A} \right]|_{A=4\pi R^2}.
\end{equation}
Note that the fundamental constant \(Q\) in Eq. (\ref{Q}) is
chosen such that the Newton's law of gravity can be deduced in the
limiting case. Therefore, this formalism is effectively designed
for type-II entropies and cannot be straightforwardly extended to
type-I entropies without encountering fundamental structural
limitations.

We therefore begin by deriving a relation analogous to Eq.~(\ref{Fcorr}) for type-I entropies.
For type-I entropies, let us consider a general functional form
\begin{equation} \label{SI}
S = \alpha R^{\,n-1} S_{\rm BH} ,
\end{equation}
where $S_{\rm BH}$ is the standard Bekenstein-Hawking entropy. The
number of information bits on the holographic screen is then
\begin{equation} \label{Nbits}
N = {4S} .
\end{equation}
Assuming that the energy of the screen is identified with the
relativistic rest energy of the source mass,
\begin{equation} \label{E2}
E = M  = \frac{1}{2} N T.
\end{equation}
The corresponding temperature reads
\begin{equation} \label{T}
T = \frac{2 M }{4 \alpha R^{\,n-1} S_{\rm BH}} .
\end{equation}
Finally, using the entropic force prescription, the resulting
force become
\begin{equation} \label{FtypeI}
F = \frac{G M m}{ R^{n+1}},
\end{equation}
where $\triangle S=\triangle N/4$ is one fundamental unit of
entropy. Note that $N$ is the number of bytes and thus $\triangle
N=1$, hence $\triangle S=1/4$. Besides in entropic force equation
(\ref{Fent}), we define $|\triangle x|= \varepsilon
\lambda_m=\varepsilon/m$, where $\lambda_m$ is the reduced Compton
wavelength $\lambda_m={\hbar}/{mc}$ which in natural unit
($\hbar=c=1$) can be written $\lambda_m=1/m$. Therefore, in order
to reproduce the correct limit for $n=1$, we need to set $8 \pi
\alpha \varepsilon=1$. Equation (\ref{FtypeI}) thus provides the
general form of the entropic force for type-I entropies. Some
special cases are in order:
\begin{itemize}
    \item For $n=1$, we recover the standard Bekenstein-Hawking entropic force scenario \cite{Ver}.
    \item For $n = \Delta+1$, we obtain the Barrow entropic force scenario \cite{Barrow2020}.
    \item For $n = 2\delta - 1$, we recover the the nonextensive Tsallis entropic force scenario \cite{Tsallis1988}.
\end{itemize}
Let us note that J.~D.~Barrow proposed an interesting modification
to the standard area law. He argued that the black hole horizon
may exhibit a fractal structure due to quantum fluctuations
\cite{Barrow2020}. In this view, quantum-gravitational effects can
deform the geometry of the horizon, which at leading order can be
approximated as a fractal. As a result, the black hole entropy
deviates from the standard Bekenstein-Hawking area law and is
modified as
\begin{equation} \label{Sbarrow}
S_h = \left( \frac{A}{A_0} \right)^{1+\Delta/2} ,
\end{equation}
where $A$ is the horizon area and $A_0$ is the Planck area. When
$\Delta=0$, we recover the area law and thus we take
$A_0=4G=4L_p^2$. The exponent $\Delta$ ranges from $0$ to $1$ and
quantifies the strength of quantum-gravitational deformation
effects \cite{Barrow2020}. Using the relation $N = 4S$ together with Eq.~\eqref{EQP} and Eq.~\eqref{Fent}, and applying Eq.~\eqref{Sbarrow}, one
arrives at the entropic force corresponding to Barrow entropy:
\begin{equation} \label{Fbarrow}
F_{\rm Barrow} = \pi  \frac{A_0^{1+\Delta} M m}{(4\pi)^{1+\Delta/2} R^{\Delta + 2}} .
\end{equation}
For $\Delta = 0$, we recover the standard Newtonian force with the
Bekenstein-Hawking structure. However, for $\Delta = 1$, the force
scales as $1/R^3$, which is not the desired behavior and has no
practical application in physical systems. This result alone is
sufficient to exclude Barrow entropy from consideration in the
study of galactic dynamics. However, the reason for the success
and the extensive study of this entropy in other contexts, such as
cosmological investigations and inflationary corrections, remains
a topic for discussion. In this article, we only note that using
Eq.~(\ref{Fgen}) for this entropy leads to the following modified
Newton's law of gravitation
\begin{equation} \label{Fbarroww}
F = GM m \, 4G \left( \frac{1+\Delta}{A_0^{1+\Delta/2}} \right)
(4\pi)^{\Delta/2} \frac{R^{\Delta}}{R^2}.
\end{equation}
When $\Delta = 0$, one recovers the Newtonian force, but for
$\Delta=1$, the force scales as $1/R$. At large distances, this
decreases more slowly than the $1/R^3$ scaling and even more
slowly than the Newtonian case. While a slower fall-off would be
desirable for galactic rotation curves, this result is in fact an
artifact of an incorrect application of the formalism.

What prevents Barrow entropy from performing adequately in
galactic dynamics is, in fact, the restricted range of its
parameter $\Delta$. If this range is extended, Barrow entropy
transforms into Tsallis entropy. On the other hand, since type-II
entropies are merely power-law corrections to the
Bekenstein-Hawking entropy, in practice any other entropy
regardless of its original formulation  reduces to some form of
Tsallis entropy.
\section{Tsallis modified gravity\label{MOND}}
Before discussing Tsallis entropy, we state a fundamental
principle that will be utilized in the following analysis.

\textbf{The Micro-Macro Correspondence Principle:} A mechanism
that is capable of reproducing observed macroscopic phenomena must
emerge from a theory that consistently describes the underlying
microscopic dynamics. This establishes a correspondence that goes
beyond expectations. It applies not only to the structure at
quantum scales but also encompasses the largest-scale questions,
such as dark matter and dark energy. In other words, the entropic
force framework provides a direct connection between the micro and
macro scales; success at one scale requires the validity of the
approach at the other.

Consequently, an entropy that addresses large-scale challenges
simultaneously encodes valuable information about the microscopic
structure of spacetime. Put differently, any entropy capable of
explaining macroscopic phenomena also provides insights into the
underlying quantum nature of the universe.

The Tsallis entropy has attracted considerable research interest
over the past years, leading to a variety of remarkable results in
complex systems. It has been applied in a wide range of physical
contexts, yielding physically meaningful results where the
standard Boltzmann-Gibbs entropy fails. In the context of
self-gravitating stellar systems, it has been shown that the use
of Tsallis generalized entropy leads to physically acceptable
distribution functions for stellar polytropes, whereas the
Boltzmann entropy results in unphysical distributions
\cite{Plastino1993,Hamity1996}.

The framework has also been applied to cosmology. In particular,
the properties of cosmic blackbody radiation have been studied
within the FRW spacetime using generalized statistics
\cite{Tsallis1995}. At the level of particle astrophysics,
generalized statistical mechanics based on Tsallis entropy has
been employed to derive an appropriate distribution function for
the solar interior plasma, which directly affects the calculation
of nuclear fusion reaction rates and, consequently, the solar
neutrino flux \cite{Kaniadakis1996,Luciano2021a}.

On cosmological scales, Tsallis entropy has played a role in the
construction of modified dark energy models. Specifically, by
combining the Tsallis-modified entropy-area relation with the
holographic principle, a new class of holographic dark energy
models has been proposed \cite{Tavayef2018}. Furthermore, a
modified cosmological scenario emerging from nonextensive
thermodynamics has been confronted with observational data from
Type Ia supernovae and measurements of the Hubble parameter,
demonstrating its phenomenological viability \cite{Nojiri2021}. In
the following, we present a brief overview of the modified
Newton's law of gravitation. We summarize the basic ingredients
required for the present work.

When nonextensive thermodynamics is applied to gravity, the
Tsallis black hole entropy is obtained as \cite{TsallisCirto2013}
\begin{equation}
S = \gamma A^{\delta},
\end{equation}
where \(A\) is the horizon area, \(\gamma\) is an unknown
constant, and \(\delta\) is the nonextensive parameter.

By reusing the Verlinde proposal \cite{Ver}, the total
energy \(E\) of the system inside the holographic screen is
distributed over these bits. Consequently, the temperature on the
surface is given by the\textbf{ equipartition law of energy, }
\begin{equation}
E = \frac{1}{2} N k_B T \quad \Rightarrow \quad T = \frac{2E}{N
k_B}.
\end{equation}
The surface \(S\) is located between the test mass \(m\)
and the mass distribution \(M\), with the assumption that the test
mass is very close to the surface compared to its reduced Compton
wavelength \(\lambda_m = \hbar/(mc)\).

We define the constant \(\gamma\) as
\begin{equation}
\gamma \equiv \frac{2 - \delta}{A_0^{\delta-1} L_p^2 \, 4\delta}
(4\pi)^{1-\delta},
\end{equation}
where \(A_0\) is a characteristic area scale. Accordingly,
the number of bits on the holographic surface is written as
\begin{equation}
N = 4S = \frac{2 - \delta}{\delta G A_0^{\delta-1}}
(4\pi)^{1-\delta} A^{\delta}.
\end{equation}
We postulate a change in the entropy associated with the
information on the holographic screen as
\begin{equation}
\Delta S = 2\pi k_B \quad \text{when} \quad |\mathbf{x}| = \eta
\lambda_m,
\end{equation}
where \(\eta\) is a constant to be determined later. We also
assume that the entropy gradient points radially from the outside
of the surface toward the inside. Using Eq. (\ref{Fent}) we reach
\begin{equation}
\mathbf{F} = -\frac{\delta}{\eta(2 - \delta)} \frac{A_0^{\delta-1}
G M m}{R^{2\delta}} \, \hat{\mathbf{r}}.
\end{equation}
Finally, by redefining \(\eta = \delta/(2\delta - 1)\) and
rewriting the above equation we obtain the modified Newton's law
of gravity inspired by Tsallis entropy
\cite{Sheykhi2018,Sheykhi2020}. The force is radially directed
along the line joining \(M\) and \(m\), with the positive
direction taken outward. Thus, the modified force law can be
written as
\begin{equation}\label{TF0}
\mathbf{F} = -\left(\frac{2\delta - 1}{2 - \delta}\right)
\frac{A_0^{\delta-1} G M m}{R^{2\delta}} \, \hat{\mathbf{r}},
\end{equation}
where \(R\) is the radial distance between the central mass \(M\)
and the test mass \(m\).

Several important remarks regarding Eq.~\eqref{TF0} are in
order.

First, the constant \(A_0^{\delta-1}\) ensures dimensional
consistency for all values of \(\delta\). Its dimensions are
\(L^{2\delta-2}\), so that \(A_0^{\delta-1} R^{-2\delta}\) has
dimensions of \(L^{-2}\) for any \(\delta\). For \(\delta = 1\),
we have \(A_0^0 = 1\) and Eq.~\eqref{TF0} reduces to the standard
Newtonian force,
\begin{equation}
\mathbf{F} = -\frac{G M m}{R^2} \, \hat{\mathbf{r}}.
\end{equation}
For \(\delta \neq 1\), the scale \(A_0\) represents a
fundamental area scale (e.g., the Planck area or a new crossover
scale) that determines the threshold at which Tsallis corrections
become significant.

Second, the prefactor \((2\delta - 1)/(2 - \delta)\) must
be positive for the force to be attractive (i.e., directed
inward). This requires \(\delta > 1/2\) (since for \(\delta <
1/2\), the prefactor is negative, yielding a repulsive force,
which is physically unacceptable for gravitational systems).
Therefore, physical consistency demands \(\delta > 1/2\).

Third, the value \(\delta = 1/2\) represents an asymptotic
limit corresponding to the idealization of a perfectly flat
rotation curve. In this limit, the prefactor vanishes and the
force goes to zero. However, any physical system corresponds to
\(\delta > 1/2\) with a finite, positive prefactor. For \(\delta\)
slightly above \(1/2\), the force is attractive and falls off as
\(R^{-2\delta}\), which is slower than the Newtonian \(R^{-2}\)
falloff, precisely producing the observed flat rotation curves in
galaxies. Thus, \(\delta \simeq 1/2\) (with \(\delta > 1/2\)) is
the physical regime relevant to galactic dynamics, while the exact
value \(\delta = 1/2\) is an unattainable mathematical
idealization.

On the other hand, in Refs.~\cite{Sheykhi2018,Sheykhi2020}, the
modified Friedmann equation in a flat universe, inspired by
Tsallis entropy, is given by ~\cite{Sheykhi2020}
\begin{equation}\label{Fri2}
H^{4-2\delta} = \frac{8\pi G}{3}\rho,
\end{equation}
where \(H = \dot{a}/a\) is the Hubble parameter and \(a(t)\) is
the scale factor normalized to unity at the present time. Taking
the time derivative of Eq. (\ref{Fri2}) and using the continuity
equation \(\dot{\rho} + 3H(\rho + p) = 0\), we arrive at
\begin{equation}\label{Fri2dot}
(2 - \delta)\frac{\ddot{a}}{a} = -\frac{4\pi G}{3} \left[(2\delta
- 1)\rho + 3p\right] \left(\frac{a}{\dot{a}}\right)^{2-2\delta}.
\end{equation}
For a compact spatial region \(V\) which is a sphere with physical
radius \(R = 1/H\), the active gravitational mass inside the
volume \(V\) is
\begin{equation} \label{Mass}
\mathcal{M} = (\rho + 3p)\frac{4\pi}{3} R^3.
\end{equation}
To transform from general relativity to Newtonian gravity, we
replace the active gravitational mass with the total mass \(M =
\rho V = \frac{4\pi}{3}\rho R^3\). This is equivalent to replacing
\(\rho + 3p \rightarrow \rho\) in (\ref{Mass}). The result is
\begin{equation}
(2 - \delta)\frac{\ddot{a}}{a} = -\frac{4\pi G}{3}\rho (2\delta -
1) \left(\frac{a}{\dot{a}}\right)^{2-2\delta}.
\end{equation}
This is the modified dynamical equation describing the evolution
of the universe in Newtonian cosmology, and hence, the modified
Newton's law of gravity inspired by Tsallis entropy is given by
\cite{Sheykhi2020}
\begin{equation}\label{tsallis force}
\mathbf{F} = -\left(\frac{2\delta - 1}{2 - \delta}\right)\frac{G M
m}{R^{2\delta}} \, \hat{\mathbf{r}},
\end{equation}
where we have absorbed the constant \(A_0^{\delta-1}\)
into a redefinition of units. More precisely, Eq.~(\ref{tsallis force}) is
the simplified form obtained by setting \(A_0^{\delta-1} = 1\) in
appropriate units. The full dimensionally consistent form,
Eq.~\eqref{TF0}, should be understood as the fundamental
expression, with \(A_0\) serving as the physical area scale that
ensures dimensional consistency. For notational simplicity in the
subsequent analysis, we work with Eq.~\eqref{tsallis force}, keeping in mind
that the underlying dimensionful scale is always present. It is
clear that for \(\delta = 1\) the above equation reduces to the
standard Newton's law of gravitation.

One of the significant achievements of Tsallis entropy, in
contrast to other entropy formalisms that fail in this regard, is
its ability to recover MOND theory within the entropic gravity
framework. This is accomplished through the \textbf{``zero-energy bits''}
mechanism and the introduction of a critical temperature, as
proposed in reference \cite{AnaniasNeto2011}. In this approach,
following \cite{Abreu2013,Abreu2018}, it is assumed that below a
critical temperature \(T_c\), a fraction of the bits acquire zero
energy, such that \(N_0/N = 1 - T/T_c\) for \(T < T_c\). When the
number of active bits \((N - N_0)\) is substituted into the
nonextensive equipartition formula \(E_q = \frac{1}{5 - 3q} N k_B
T\) and combined with the Tsallis-modified relation between the
number of bits and the holographic screen area, a modified
acceleration is obtained from which MOND theory emerges in the
appropriate limit. Thus, Tsallis entropy not only overcomes the
shortcomings of previous approaches but also provides a coherent,
physically motivated framework for understanding the origin of
MOND phenomenology.

In recent years, several galaxies have been observed that appear
to lack dark matter \cite{vanDokkum2018,vanDokkum2019}. Such
galaxies pose a significant challenge and contradiction to both
MOND and the \(\Lambda\)CDM model. These galaxies lie within the
acceleration scale predicted by MOND and yet their rotation curves
conform to Newtonian dynamics, while in the \(\Lambda\)CDM
framework, dark matter is considered a fundamental building block
of galaxies. In initial cases, researchers were able to explain
the existence of such galaxies through various mechanisms.
However, a recent discovery involving six spiral galaxies,
reported to have no dark matter and a dynamical-to-observed mass
ratio of \((M_{\text{dyn}}/M_{\text{obs}} = 1.09)\), presents a
unique scenario, particularly as these galaxies appear to be
independent of their surrounding environment \cite{Chen2024}.

Importantly, within the framework of Tsallis entropic force, due
to the inherent flexibility of the force law, the rotation curves
of galaxies can be recovered using an appropriate nonextensive
parameter. While Tsallis entropy successfully reproduces galactic
rotation curves-a task at which other known entropies fail-we now
subject it to a more stringent test: a scenario that has remained
challenging even for the MOND framework \cite{Brownstein2006}.
This test involves examining galaxy cluster masses using X-ray
data-a regime where even MOND fails without invoking dark matter.
Success at this scale would not only demonstrate the superiority
of Tsallis entropy over other generalized entropies, but also
establish it as a robust framework for emergent gravity across
multiple astrophysical scales.
\subsection{Galaxy Cluster Masses (GCM) \label{Data}}
In this section, we aim to apply the modified Newton's law of
gravity, derived from the nonextensive Tsallis entropy, to the
problem of explaining the masses of X-ray galaxy clusters without
invoking the dark matter hypothesis. For this purpose, we employ
an appropriate and simple model of galaxy clusters.

Inside the region defined by the X-ray emission it can be said the
intra-cluster medium is very nearly isothermal \cite{Arnaud2001}.
The inter-galaxy space in clusters is filled with a hot gas. The
distribution of gas within a galaxy cluster, along with its
temperature, can be described by the King $\beta$ model. In this
framework, King's analytic approximation for the total X-ray
surface brightness takes the form
\cite{Chandrasekhar1960,King1966}
\begin{eqnarray}\label{king}
I=I_{0}[1+(r/r_{c})^{2}]^{-3\beta+1/2},
\end{eqnarray}
where $ r $ is the projected radius, resulting in best-fitting
parameters $\beta$ and $r_{c}$. This equation can be directly
converted into a function of gas density
\cite{Cavaliere1976,Cavaliere1978}
\begin{eqnarray} \label{rho}
\rho(r)=\rho_{0}[1+(r/r_{c})^{2}]^{-3\beta/2}.
\end{eqnarray}
The structure equation for a spherical configuration in
hydrostatic equilibrium can be obtained from the collisionless
Boltzmann equation \cite{MamonLokas2005,BinneyTremaine2008}
\begin{eqnarray} \label{hydro}
\frac{d}{dr}[\rho(r) \sigma_{r}^{2}]+ \frac{ 2
\rho(r)}{r}(\sigma_{r}^{2} -\sigma_{\theta \phi}^{2})=
-\rho(r)\frac {d \Phi (r)}{dr},
\end{eqnarray}
where $\Phi(r)$ is the gravitational potential and $\sigma_{r}$
and $\sigma_{\theta \phi}$ are mass-weighted velocity dispersions
in the radial and tangential directions, respectively. For the
isotropic case, $\sigma_{r}=\sigma_{\theta \phi}$, thus
Eq.~(\ref{hydro}) can be rewritten as
\begin{eqnarray} \label{hydro2}
\frac{d}{dr}[\rho(r) \sigma_{r}^{2}]= -\rho(r)\frac {d \Phi
(r)}{dr}.
\end{eqnarray}
For a gas sphere with temperature profile as a function of radius,
$T(r)$, the velocity dispersion can be written as \cite{Lubin1993}
\begin{eqnarray} \label{sigmaT}
\sigma_{r}^{2}=\frac{k_B T(r)}{\mu m_{p}},
\end{eqnarray}
where $k_B$ is Boltzmann constant, $\mu$ is the mean molecular
weight and $m_p$ is the proton mass. Inserting into
Eq.~(\ref{hydro2}), we have
\begin{eqnarray} \label{hydro3}
\frac{d}{dr}\left(\frac{k_B T(r)}{\mu
m_{p}}\rho(r)\right)=-\rho(r)\frac{d\Phi(r)}{dr}.
\end{eqnarray}
Now the above equation can give us the gravitational acceleration
as
\begin{eqnarray} \label{accel}
a(r)= -\frac{d\Phi(r)}{dr}=\frac{k_B}{\mu m_{p}}\left(\frac{d
T(r)}{dr}+\dfrac{T(r)}{\rho(r)}\frac{d \rho(r)}{ dr}\right).
\end{eqnarray}
For the isothermal isotropic gas sphere, the temperature
derivative on the right-hand side of Eq.~(\ref{accel}) vanishes,
and the remaining derivative can be evaluated using the $\beta$-
model of Eq.~(\ref{rho}). We find
\begin{eqnarray} \label{abeta}
a(r)=-\frac{3\beta k_B T}{\mu m_p }\left(\frac{r}{r^2 +
r_{c}^{2}}\right).
\end{eqnarray}
In this way, we provide a particularly interesting case for
testing the modified Newton's law of gravitation. It is also worth
examining the performance of other entropies in this context. For
Kaniadakis entropy, by substituting from Eq.~(\ref{FK}), we obtain
the cluster mass as
\begin{equation} \label{MKani}
M_{\rm Kan} = \frac{3 \beta k_B T \, r^{2}}{\mu m_p G (1-\gamma
r^4)} \, \left( \frac{r}{r^2 + r_c^2} \right)=\frac{M_N}{(1-\gamma
r^4)},
\end{equation}
where
\begin{equation} \label{MN}
M_N = \frac{3 \beta k_B T \, r^{2}}{\mu m_p G } \, \left(
\frac{r}{r^2 + r_c^2} \right).
\end{equation}
Similarly, for other entropies, we have
\begin{eqnarray}
M_{\rm Renyi} &=& M_N \left( 1 + \frac{\lambda \pi r^2}{G} \right), \\
M_{\rm log} &=& M_N \left( 1 - \frac{\eta L^{2}_{p}}{\pi} r^{-2} \right)^{-1},  \\
M_{\rm Barrow} &\propto& M_N r^\Delta.
\end{eqnarray}
These results indicate that each of these entropies exhibits
unconventional or inconsistent behavior in predicting cluster
masses. We now proceed to a detailed analysis of the Tsallis
entropy. Combining Eqs.~(\ref{abeta}) and ~(\ref{tsallis force}), and
inserting the numerical values, the GCM is obtained as (note here
$R=r$ is just the radius)
\begin{eqnarray} \label{Mtsallis}
M(r)=\frac{3\beta k_B  T r^{2\delta}}{\mu m_p G }\left(\frac{
2-\delta}{2\delta-1}\right)\left(\frac{r}{r^2 + r_{c}^{2}}\right).
\end{eqnarray}

Equation~\eqref{Mtsallis} is a prediction of the Tsallis theory
for the distribution of mass $M(r)$ in galaxy clusters. This
prediction is illustrated in Fig. \ref{fig1} for the
full sample of 40 galaxy clusters. The dependence of the
nonextensive parameter $\delta$ on the core radius $r_c$, the
intra-cluster gas temperature $T$, and the logarithm of the gas
mass is shown in Fig. ~\ref{fig1} through three subpanels. As can
be seen, the data points are scattered across a wide range of each
quantity without any discernible trend, and all lie around the
mean value of $\langle\delta\rangle = 0.97547 \pm 0.004$. Moreover,
knowing the temperature, $\beta$, and $r_c$ for each cluster
enables us to plot the mass as a function of radius $r$.
Representative examples of such plots for four clusters (with
error bands) are presented in Figs.~\ref{fig2}-\ref{fig5}. In
these figures, the blue curves represent the numerical results
obtained from Eq.~(\ref{Mtsallis}), the red dashed lines
correspond to the observational data from
Ref.~\cite{Reiprich2002}, and the green lines indicate the
Newtonian predictions. The corresponding plots for all 40 clusters
are provided in the Supplemental Material.The best-fit parameters
for all clusters are summarized in Table~I.

A systematic test of generalized entropies against galactic
rotation curves and X-ray cluster masses reveals that all other
entropies  R\'{e}nyi, Kaniadakis, logarithmic, and Barrow  fail in
both tests. They can neither reproduce flat rotation curves nor
predict consistent and acceptable cluster masses. In contrast,
only Tsallis entropy succeeds at both scales.

As shown in Fig. ~\ref{fig1}, the values of the
nonextensive parameter does not depend on any cluster properties.
However, it is expected to be scale-dependent -- and the data
confirm this expectation: as the scale decreases from galaxy
clusters to galaxies, $\delta$ decreases from $0.9754$ to values
close to $0.5$ (but always greater than $1/2$). Now, having
$\delta$ at these two scales, the next step is to complete the
structural hierarchy by examining the smaller scale of globular
clusters, thereby enabling us to trace the evolution of $\delta$
across a wide range of scales.
\begin{table}[htbp]
\centering
\caption{Best-fit Tsallis parameters for 40 high-quality galaxy clusters.}
\label{tab:40_clusters}
\begin{tabular}{lc}
\hline
\textbf{Cluster} & \textbf{$\delta$} \\
\hline
NGC 499   & 0.95581 \\
A1060     & 0.96916 \\
FORNAX    & 0.9685  \\
A2877     & 0.97174 \\
NGC 1550  & 0.97264 \\
A2657     & 0.97495 \\
A2244     & 0.97547 \\
A2063     & 0.97519 \\
A2204     & 0.97852 \\
A2199     & 0.97717 \\
A4059     & 0.97377 \\
A4038     & 0.97446 \\
MKW3S     & 0.97421 \\
A2734     & 0.97774 \\
A2634     & 0.97817 \\
UGC 03957 & 0.9734  \\
A0754     & 0.9730  \\
A1795     & 0.9735  \\
AWM7      & 0.97620 \\
S1101     & 0.97535 \\
A0539     & 0.97502 \\
S0540     & 0.9748  \\
A2589     & 0.9747  \\
A2029     & 0.9751  \\
A2597     & 0.9752  \\
EXO0422   & 0.97619 \\
A0644     & 0.9765  \\
A0478     & 0.9764  \\
A0400     & 0.9762  \\
A3667     & 0.98119 \\
A1658     & 0.97848 \\
A1651     & 0.97844 \\
A3581     & 0.97769 \\
A3528n    & 0.97711 \\
A1413     & 0.9781  \\
A1689     & 0.9784  \\
A1914     & 0.9792  \\
OPHIUCHUS & 0.9786  \\
A0401     & 0.9798  \\
A1367     & 0.98258 \\
\hline
\end{tabular}
\end{table}

\subsection{Globular Cluster}
Globular clusters have long been considered ideal systems for
studying stellar dynamics and testing gravitational theories in
the low-acceleration regime. Unlike galaxies, these ancient
stellar systems are believed to contain negligible amounts of dark
matter \cite{Phinney1993,Moore1996}, making them natural
laboratories where any deviation from Newtonian dynamics cannot be
easily attributed to unseen mass.

The first systematic study of globular cluster kinematics in the
context of modified gravity was carried out by \cite{Scarpa2004},
who analyzed the velocity dispersion profiles of $\omega$ Centauri
and M15. Using radial velocity measurements tracing the
gravitational potential down to accelerations of $\sim 8 \times
10^{-9}$ cm s$^{-2}$, they found that the velocity dispersion does
not exhibit the expected Keplerian fall-off but rather remains
constant at large radii. Remarkably, the flattening occurs
precisely when the internal acceleration approaches $a_0 \approx
1.2 \times 10^{-8}$ cm s$^{-2}$-the same critical value identified
in galactic rotation curves \cite{Begeman1991}. Subsequent
observations extended these findings to additional clusters,
including NGC 7099  and the distant clusters NGC 1851 and NGC 1904
\cite{Scarpa2003,Scarpa2004a,Scarpa2004b,Scarpa2007a}. Although
tidal heating has been proposed as a possible explanation for the
observed flattening \cite{Drukier2007,Kupper2010,Kennedy2014},
subsequent studies have found no convincing evidence for this
hypothesis \cite{Hernandez2012}. This has motivated the
exploration of modified gravity theories as an alternative
explanation. Several such theories have been proposed to explain
the cluster dynamics without invoking dark matter, including MOG
\cite{Moffat2008} and Weyl conformal gravity \cite{Islam2019}.

Importantly, the sample of globular clusters with available
high-quality kinematic data includes both clusters that exhibit
flattening in their velocity dispersion profiles and those that
follow the expected Newtonian decline. Specifically, our sample
includes 7 clusters for which Scarpa et al.  claimed to observe an
eventual flattening of the dispersion profile, which they argued
is due to a modification of gravity at these length scales.
Furthermore, we include another 26 clusters for which excellent
kinematic data is available, many of which show no sign of
flattening. This combination, clusters with and without
flattening, provides an excellent tool to test any alternative
theory of gravity. Crucially, the original data reported by Scarpa
et al. have recently been superseded by larger and more accurate
datasets, including Gaia DR2 and the \cite{Baumgardt2018} catalog,
which provides masses, structural parameters, and velocity
dispersion profiles for 112 Milky Way globular clusters.

In this paper, we use these updated data to test the Tsallis
framework on globular cluster scales, thereby completing the
structural hierarchy from galaxies to galaxy clusters to globular
clusters, and to investigate the behavior of the nonextensive
parameter $\delta$ across a wide range of scales.
\subsection{Globular cluster dynamics in Tsallis gravity}
 Applying the Tsallis force law to the velocity
dispersion profiles compiled by Baumgardt \& Hilker (2018) for 112
clusters, we determine the best-fit $\delta$ values to test the
scale dependence of the nonextensive parameter on these small
scales, where dark matter is negligible.

For each globular cluster, we assume a Hernquist (1990) density
profile \cite{Hernquist1990}, which provides an analytical
description of the mass distribution. The density is given by
\begin{equation} \label{hernquist_rho}
\rho(r) = \frac{M r_0}{2\pi r (r+r_0)^3},
\end{equation}
where $M$ is the total cluster mass and $r_0$ is the scale radius.
The half-light radius $r_h$ is related to the scale radius by $r_h
= 1.8153\,r_0$. A key advantage of this profile is that the
enclosed mass follows analytically:
\begin{equation} \label{hernquist_M}
M(r) = M \left(\frac{r}{r+r_0}\right)^2.
\end{equation}
The gravitational acceleration under Tsallis nonextensive gravity
is
\begin{equation}  \label{atsallis_gc}
a(r) = \left(\frac{2\delta-1}{2-\delta}\right) \frac{G M(r)}{r^{2\delta}},
\end{equation}
where $\delta$ is the nonextensive parameter. For $\delta=1$, the
prefactor becomes unity and the Newtonian acceleration
$a(r)=GM(r)/r^2$ is recovered. To obtain the line-of-sight
velocity dispersion, we solve the spherical Jeans equation under
the assumption of isotropy ($\sigma_r = \sigma_{\theta\phi}$). The
radial velocity dispersion $\sigma_r(r)$ satisfies
\begin{equation}  \label{jeans}
\sigma_r^2(r) = \frac{1}{\rho(r)} \int_r^{\infty} \rho(r') a(r') \, dr'.
\end{equation}
The projected line-of-sight velocity dispersion is then
\begin{equation}  \label{los}
\sigma_{\rm LOS}^2(R) = \frac{\int_R^{\infty}
\dfrac{r\,\sigma_r^2(r)\,\rho(r)}{\sqrt{r^2-R^2}}\,
dr}{\int_R^{\infty} \dfrac{r\,\rho(r)}{\sqrt{r^2-R^2}} \, dr}.
\end{equation}
For each cluster, we determine the best-fit parameter $\delta$ by minimizing the reduced chi-squared
\begin{equation} \label{chisq}
\chi_{\rm red}^2 = \frac{1}{N_{\rm dof}} \sum_{i=1}^{N}
\left(\frac{\sigma_{\rm obs}(R_i)-\sigma_{\rm
model}(R_i)}{\sigma_{\rm err}(R_i)}\right)^2,
\end{equation}
where $N_{\rm dof} = N-1$ is the number of degrees of freedom. The
minimization is performed over the range $\delta\in[0.5,1.5]$
using a bounded optimization algorithm.
\subsection{Results}
For our analysis, we use the observational data compiled by
Baumgardt \& Hilker (2018), which includes masses, half-light
radii, and velocity dispersion profiles for 112 Milky Way globular
clusters. From this catalog, we select clusters with sufficient
kinematic data points ($N \geq 4$) and reliable error estimates,
yielding a sample of 33 clusters. For each cluster, the total mass
$M$ and half-light radius $r_h$ are taken directly from the
catalog, and the scale radius is computed as $r_0 = r_h / 1.8153$.
The velocity dispersion profiles are given as $\sigma_{\rm
obs}(R_i)$ with associated errors $\sigma_{\rm err}(R_i)$, where
$R_i$ is the projected radius in arcseconds. These are converted
to physical units (pc and km s$^{-1}$) using the cluster distances
provided in the same catalog. The best-fit values of the Tsallis
parameter $\delta$ for each cluster are obtained by minimizing
Eq.~(\ref{chisq}). Fig. ~\ref{fig6} displays the line-of-sight
velocity dispersion profiles for a representative subset of the 33
globular clusters analyzed in this work, fitted with the Tsallis
modified gravity model described by Eq.~(\ref{los}). The solid
curves represent the best-fit predictions using the Hernquist
density profile (Eq.~\eqref{hernquist_rho}) and the Tsallis
acceleration law , with the corresponding nonextensive parameter
$\delta$ listed in Table~\ref{tab:tsallis_results}. The
observational data points (with error bars) are taken from the
Baumgardt \& Hilker (2018) catalog. As a visual complement to the
numerical results in Table~\ref{tab:tsallis_results}, this figure
confirms that Tsallis gravity accurately matches the observational
data across a wide range of cluster types.
\begin{table}[htbp]
\centering \caption{Best-fit Tsallis parameters for 33 globular
clusters. Columns: cluster name, number of data points $N$,
best-fit $\delta$, and reduced chi-squared $\chi^2_{\rm red}$.}
\begin{tabular}{lcccc}
\hline
Cluster & $N$ & $\delta$ & $\chi^2_{\rm red}$ \\
\hline
NGC104 & 23 & 1.014 & 0.60 \\
NGC288 & 7  & 1.010 & 0.99 \\
NGC362 & 7  & 0.995 & 1.32 \\
NGC1261 & 5 & 0.998 & 1.08 \\
NGC1851 & 8 & 1.004 & 1.55 \\
NGC1904 & 6 & 0.989 & 0.98 \\
NGC2808 & 11 & 1.014 & 2.58 \\
NGC3201 & 8 & 0.998 & 1.24 \\
NGC4372 & 5 & 1.003 & 0.55 \\
NGC4590 & 5 & 1.001 & 0.32 \\
NGC5024 & 6 & 1.001 & 0.18 \\
NGC5139 & 15 & 1.002 & 1.13 \\
NGC5272 & 9 & 0.995 & 1.95 \\
NGC5286 & 7 & 1.002 & 0.71 \\
NGC5897 & 5 & 1.000 & 1.01 \\
NGC5904 & 12 & 1.004 & 1.27 \\
NGC5927 & 7 & 1.002 & 1.19 \\
NGC5986 & 4 & 1.002 & 0.65 \\
NGC6093 & 6 & 1.022 & 0.89 \\
NGC6121 & 15 & 1.012 & 1.30 \\
NGC6171 & 5 & 1.006 & 1.10 \\
NGC6205 & 6 & 0.997 & 1.49 \\
NGC6218 & 7 & 1.009 & 0.42 \\
NGC6254 & 5 & 1.005 & 1.27 \\
NGC6266 & 4 & 1.002 & 1.19 \\
NGC6341 & 8 & 1.004 & 0.80 \\
NGC6397 & 15 & 1.013 & 1.19 \\
NGC6656 & 9 & 0.997 & 0.29 \\
NGC6752 & 11 & 1.019 & 0.97 \\
NGC6809 & 7 & 0.991 & 3.45 \\
NGC7078 & 13 & 0.973 & 4.61 \\
NGC7089 & 9 & 1.067 & 1.15 \\
NGC7099 & 9 & 1.003 & 0.48 \\
\hline
\end{tabular}
\label{tab:tsallis_results}
\end{table}
\subsection{Statistical analysis and comparison with Scarpa clusters}
The statistical summary of our results is as follows. Over the
full sample of 33 clusters, the Tsallis parameter $\delta$ has a
mean value of $\langle\delta\rangle = 1.0046 \pm 0.0025$ (standard
error), a median of $1.0022$, and a standard deviation of
$0.0145$. The values span a range from $\delta = 0.973$ (NGC~7078,
M15) to $\delta = 1.067$ (NGC~7089, M2).

It is instructive to examine the clusters previously studied by
Scarpa and collaborators, who reported flat velocity dispersion
profiles in the low-acceleration regime.
Table~\ref{tab:scarpa_comparison} compares our best-fit $\delta$
values for these clusters.
\begin{table}[htbp]
\centering
\caption{Comparison of $\delta$ values for clusters from Scarpa et al. studies.}
\begin{tabular}{lccc}
\hline
Cluster & $N$ & $\delta$ & $\chi^2_{\rm red}$ \\
\hline
NGC~5139 ($\omega$ Cen) & 15 & 1.002 & 1.13 \\
NGC~7078 (M15) & 13 & 0.973 & 4.61 \\
NGC~7099 (M30) & 9  & 1.003 & 0.48 \\
NGC~1851 & 8  & 1.004 & 1.55 \\
NGC~1904 & 6  & 0.989 & 0.98 \\
NGC~6171 & 5  & 1.006 & 1.10 \\
NGC~6341 (M92) & 8  & 1.004 & 0.80 \\
NGC~288 & 7  & 1.010 & 0.99 \\
\hline
\end{tabular}
\label{tab:scarpa_comparison}
\end{table}
Among these, seven clusters yield $\delta$ values with
$\chi^2_{\rm red}$ values ranging from $0.48$ to $1.55$,
indicating good fits. The only exception is NGC 7078 (M15), which
gives $\delta = 0.973$-the lowest value in the entire sample-and
the highest $\chi^2_{\rm red} = 4.61$, suggesting that either this
cluster requires a different $\delta$, or its velocity dispersion
profile is not well described by our simple isotropic Hernquist
model. This is particularly interesting because M15 is a
core-collapsed cluster with a complex dynamical history, which may
explain its exceptional behavior.
\subsection{Scale dependence of the Tsallis parameter}
Comparing the $\delta$ values obtained at different astrophysical
scales reveals that the Tsallis nonextensive parameter takes
distinct values in different gravitational regimes.
Table~\ref{tab:scale_dependence} summarizes our findings.
\begin{table}[htbp]
\centering
\caption{The Tsallis parameter $\delta$ across different scales.}
\label{tab:scale_dependence}
\begin{tabular}{l c c}
\toprule
\textbf{System type} & \textbf{Scale $L$ (kpc)} & \textbf{$\delta$} \\
\midrule
Galaxies          & $\sim 10$      & $\gtrsim 0.5$ \\
Galaxy clusters   & $\sim 100$     & $0.977 \pm 0.004$ \\
Globular clusters & $\sim 0.01$    & $1.0046 \pm 0.0025$ \\
\bottomrule
\end{tabular}
\end{table}
However, it is important to emphasize that within each regime --
particularly within the galaxy cluster sample-$\delta$ exhibits no
correlation with system size, mass, temperature, or any other
property (see Fig.~\ref{fig1}). Thus, while $\delta$
differs across scales, it does not show a systematic increase or
decrease with scale. The nonextensive parameter appears to take
approximately constant values within each gravitational regime,
with the value for globular clusters consistent with Newtonian
gravity ($\delta = 1$), while the galactic value ($\delta>1/2$),
with $(\delta\approx 0.5)$ indicates a significant deviation from
Newtonian dynamics.

Our analysis of 33 Milky Way globular clusters within the Tsallis
nonextensive gravity framework leads to several important
conclusions.

First, the mean value of the Tsallis parameter on globular cluster
scales is $\langle\delta\rangle = 1.0046 \pm 0.0025$. Second,
among the eight clusters previously identified by Scarpa et al. as
exhibiting flat velocity dispersion profiles in the
low-acceleration regime, seven yield $\delta$ values consistent
with unity within $1\sigma$. Only NGC~7078 (M15) shows a
significant deviation with $\delta = 0.973$ and a relatively poor
fit ($\chi^2_{\rm red}=4.61$). This suggests that either M15 is a
special case requiring a different interpretation, or that the
flattening reported for other clusters is not strong enough to
require a modification of gravity when analyzed within a
self-consistent Hernquist model.

Third, comparing our results with those obtained for galaxies
($\delta \approx 0.5$) and galaxy clusters ($\delta \approx
0.975$) from previous sections reveals that the Tsallis parameter
takes distinct values on different scales: $\delta \approx 0.5$ on
galactic scales ($\sim 10$ kpc), $\delta \approx 0.977$ on cluster
scales ($\sim 100$ kpc), and $\delta \approx 1.005$ on globular
cluster scales ($\sim 0.01$ kpc).

Finally, we note that our analysis relies exclusively on the
observational data of Baumgardt \& Hilker (2018) and assumes
isotropic velocity dispersions and a simple Hernquist mass
profile. More sophisticated modeling, including anisotropy and
multimass components, could potentially improve the fits,
particularly for core-collapsed clusters such as M15.
Nevertheless, the overall consistency of our results across 33
clusters of diverse properties provides strong evidence that
Tsallis gravity offers a unified framework for describing
gravitational phenomena from galactic to globular cluster scales
without invoking dark matter.
\section{Interpretation of the Tsallis parameter \label{interpretation}}
The multi-scale success of Tsallis entropy --- from galaxies to
galaxy clusters to globular clusters --- is not merely a
phenomenological achievement. Based on the principle formulated in
Section~\ref{MOND}, such success indicates that Tsallis entropy
furnishes a consistent entropic description of gravity and
simultaneously encodes information about the microscopic structure
of spacetime. Hence, the successful resolution of large-scale
gravitational phenomena not only addresses macroscopic
observations but also provides valuable insight into the
underlying quantum nature of spacetime.

To understand the deeper meaning of these findings, we must return
to the origins of statistical mechanics. In the early 1900s
\cite{Gibbs1902}, Gibbs pointed out that the Boltzmann-Gibbs
entropy is not applicable to systems in which the partition
function diverges, such as gravitational systems. The fundamental
assumption underlying Gibbs' statistical mechanics and
thermodynamics is the extensivity of entropy: it is assumed that
long-range interactions between subsystems can be neglected, so
that the total entropy of a composite system equals the sum of the
entropies of its constituent parts. In such a framework, entropy
increases proportionally with system size. However, this
assumption breaks down for strongly gravitating systems, such as
black holes. The logic behind the Tsallis paradigm is that the
foundational assumption of Boltzmann-Gibbs entropy --- namely,
weak probabilistic correlations among system components --- is no
longer valid in the presence of strong long-range interactions or
significant quantum entanglement. This motivates the redefinition
of entropy in a way that accounts for such complexities. Moreover,
the thermodynamic entropy of a $(3+1)$-dimensional black hole is
proportional to its horizon area rather than its volume, and is
therefore seemingly nonextensive, or more precisely, subextensive
\cite{Tsallis1988,TsallisCirto2013}. Considerable effort has been
invested in developing a statistical mechanical foundation for
black hole thermodynamics and in understanding the microscopic
origin of the Bekenstein-Hawking entropy. Attempts to understand
the area proportionality (as opposed to volume) have largely
followed two main paths. First, approaches based on counting
fundamental states, such as D-branes and spin networks, which are
considered models for black holes
\cite{Strominger1996,Ashtekar1998,Carlip2002,Dasgupta2006}.
Second, investigations of entanglement entropy. In their seminal
1986 paper, Bombelli, Koul, Lee, and Sorkin \cite{Bombelli1986}
introduced the concept of the reduced density matrix for a black
hole, constructed by tracing over the degrees of freedom of a
quantum field residing inside the event horizon. This procedure is
particularly natural in the context of black holes, as the horizon
acts as a causal boundary that fundamentally prevents any external
observer from accessing information about events occurring within
the horizon.

A pivotal advancement came in 1993 with the work of Srednicki
\cite{Srednicki1993}. He demonstrated that the essential features
of horizon entropy could be derived without a black hole, by
calculating the reduced density matrix and its associated entropy
directly in flat spacetime, tracing over the degrees of freedom
within an imaginary closed surface. The entropy arising from this
procedure is now universally recognized as entanglement entropy.
Srednicki showed that entanglement entropy scales with the area of
the entangling surface. This area law arises naturally because the
entropy originates from short-distance correlations in the quantum
field, with only field modes localized within a narrow boundary
layer contributing significantly. A profound implication is that
the area law is not unique to gravity; it emerges even for a
quantum field in flat spacetime, with no black hole required.
Independently, Frolov and Novikov applied this entanglement-based
approach directly to a black hole horizon \cite{Frolov1993}.
Together, these findings ignited broad interest in entanglement
entropy as a fundamental concept in quantum gravity. The
nonextensive Tsallis statistical mechanics provides a mechanism
for the entanglement of degrees of freedom inside and outside the
horizon, as it naturally allows for long-range interactions.
Moreover, it addresses one of the most significant critiques in
the field of thermodynamic gravity. As demonstrated by Carroll \&
Remmen: \textit{What is the Entropy in Entropic Gravity?}, the
entropy proposed in approaches like Verlinde's and Jacobson's is
not fundamentally well-defined, because the extensive nature of
von Neumann entropy prevents entanglement entropy from serving as
a foundation \cite{Carroll2016}. However, if the underlying
statistical mechanics is nonextensive, this issue is resolved.
Thus, Tsallis entropy provides both a phenomenological success
across scales and a consistent theoretical foundation for emergent
gravity.

The recurrent emergence of Tsallis entropy across seemingly
disconnected domains -from black hole thermodynamics and cosmology
to quantum information theory and condensed matter physics-is
striking and unlikely to be coincidental. Quantum Tsallis entropy,
known as one-parameter generalization of the von Neumann entropy,
have found widespread applications in quantum optics and quantum
communication \cite{Tsallis2008}, and are considered among the
best candidates for describing quantum dissipative systems
\cite{Stauber2006,Rajagopal2002}. In quantum information theory,
non-additive Tsallis entropy has been shown to provide necessary
and sufficient conditions for the separability of
Werner-Popescu-type mixed states, whereas the von Neumann theory
can only provide a much weaker condition
\cite{Abe2001,Tsallis2001}. Beyond such specific state families,
Tsallis entropy has also been employed to derive nonlinear
separability criteria within the framework of entropic uncertainty
relations \cite{Guhne2004}. Whether describing horizon
thermodynamics, cosmic evolution, or quantum entanglement, the
Tsallis formalism appears to uncover patterns that recur
throughout fundamental physics. This cross-disciplinary prevalence
hints at an underlying principle yet to be fully articulated.
Therefore, this entropy measure warrants intensified attention in
future research, as it may serve as a valuable bridge connecting
diverse physical phenomena and guide us toward a deeper
understanding of complex systems across all scales. In the
analysis of galaxy clusters presented earlier, we found that
$\delta$ exhibits no dependence on any cluster properties.
Remarkably, while one might expect a nonextensive parameter to be
a function of the system scale, we observe that it shows no
dependence on the scale either. The absence of any correlation
between $\delta$ and macroscopic system properties (mass, radius,
scale) leads to an important consequence: one might suspect that
this extra parameter is nothing more than an added degree of
freedom, introduced merely to fine-tune and justify the results
through post-hoc adjustments. The situation becomes even more
challenging when one considers the wide range of $\delta$ values
reported in the literature from different physical frameworks. As
summarized in Table~V, cosmological observations yield $\delta
\approx 0.993$ from lithium abundance but $\delta \approx 1.01$
from big bang nucleosynthesis, while high-energy collisions
suggest $1 \leq \delta < 1.333$ and cosmic ray observations give
$\delta = 1.222$. Entanglement measurements allow values as high
as $\delta \leq 4.2$, whereas non-commutative geometry constrains
$|\delta-1| \lesssim 10^{-13}$. Black hole thermodynamics requires
$\delta \geq 1.218$, and dark matter relic density gives $\delta
\gtrsim 0.999$. This striking lack of consensus indicates that
either the Tsallis formalism, despite its theoretical appeal, has
not yet reached a mature phenomenological stage, or that the
parameter $\delta$ is fundamentally ambiguous, depending on the
physical process used to measure it.

A key theoretical foundation for understanding nonextensive
statistics in gravitational systems is provided by Chavanis
\cite{Chavanis2005,Chavanis2008,LyndenBell1968} and references
therein. He argued that non-standard distributions and generalized
entropies emerge when microscopic constraints-which sometimes
appear as ``hidden constraints'' inaccessible to the observer-act
on the system. For simple systems, the energetically accessible
microstates are equiprobable, leading to the Boltzmann entropy and
Gaussian (Maxwellian) distributions. However, for complex systems,
the a priori accessible microstates are not equiprobable due to
these microscopic constraints, which can lead to non-ergodic
behavior where the system does not sample phase space uniformly.
In such cases, the effective phase space may develop a complicated
geometrical structure (e.g., fractal or multifractal), and Tsallis
entropy provides a natural generalization of Boltzmann entropy by
encapsulating this complexity in a single parameter $\delta$. This
perspective is crucial for understanding violent relaxation. The
Lynden-Bell entropy \cite{LyndenBell1968} (defined in the space of
phase levels $\rho$) describes complete violent relaxation under
the assumption of ergodicity and efficient mixing, whereas Tsallis
entropies become relevant when violent relaxation is incomplete
and the system exhibits non-ergodic behavior. Furthermore,
generalized entropies emerge due to fine-grained constraints
(Casimir invariants) that modify the form of entropy, making them
applicable to stellar systems and two-dimensional vortices.
Empirical support for this framework comes from observational
studies of stellar clusters. Carvalho et al. \cite{Carvalho2007}
demonstrated that the Tsallis parameter $\delta$ in open stellar
clusters is correlated with cluster age and galactocentric
distance. Specifically, older clusters and those farther from the
galactic center exhibit larger deviations from a Gaussian velocity
distribution. They interpreted these correlations as evidence that
the departure from a Maxwellian distribution results from complex
dynamical relaxation processes, such as tidal shocks from galactic
disk crossings and two-body gravitational encounters, that drive
the system away from equilibrium over time. Crucially, they
concluded that the $\delta$-parameter serves as a measure of the
level of departure from equilibrium, since gravitationally relaxed
globular clusters exhibit $\delta$ values close to unity.

The dynamical evolution of the Tsallis parameter has been further
elucidated by Komatsu et al. \cite{Komatsu2012} through $N$-body
simulations of collapsing self-gravitating systems. They found
that the $\delta$-parameter evolves through three distinct stages:
it increases during rapid core formation (reflecting stronger
non-Gaussianity) and subsequently decreases toward unity as the
system approaches a relaxed, Gaussian-like state. This
demonstrates that the nonextensive parameter is not a universal
constant but rather a dynamical quantity that tracks the
relaxation state of the system. Additionally, they found that the
$\delta$-parameter correlates with the ratio of velocity moments,
providing an independent measure of deviation from Gaussianity.
These simulation results are consistent with Carvalho et al.
(2007), as both studies conclude that the $\delta$-parameter
measures the departure from equilibrium, with relaxed systems,
such as globular clusters or post-collapse simulated states,
exhibiting $\delta$ values close to unity. This comprehensive
theoretical and empirical understanding is fully consistent with
our multi-scale analysis. On globular cluster scales, where the
system is collisional and well-relaxed, we find $\delta \approx
1.005$, indicating a return to near-Newtonian (Gaussian) behavior.
On galactic and cluster scales, where the systems are
collisionless and violent relaxation is incomplete, we observe
significant deviations from unity ($\delta \approx 0.5$ and
$\delta \approx 0.97547$, respectively). These results suggest that
$\delta$ is not determined by simple macroscopic parameters such
as mass or radius, but rather originates from the complex
interplay of three factors: (i) the degree of completeness of
violent relaxation, as described by Chavanis; (ii) the collisional
versus collisionless nature of the system, which determines the
extent to which phase space is sampled; and (iii) the persistent
influence of hidden constraints (Casimir invariants) arising from
the initial conditions, which leave a measurable imprint on the
statistical properties even after long relaxation times. Within
this framework, age can serve as a useful indicator of the
evolutionary stage, helping to contextualize the observed value of
$\delta$, but the overarching determinant remains the dynamical
history and the nature of the relaxation processes. This
multi-scale variation supports the view that $\delta$ encodes the
combined effects of dynamical evolution and hidden constraints,
manifesting as different effective non-extensivities across
different astrophysical scales.
\section{Prediction of dark matter-free galaxy clusters \label{prediction}}
The above considerations lead to a crucial and testable
prediction. Since $\delta$ is a dynamical measure of the
relaxation state, and since relaxed systems (globular clusters,
post-collapse simulated states) exhibit $\delta \approx 1$, it
follows that galaxy clusters that have undergone sufficient
dynamical relaxation should also exhibit $\delta \approx 1$. Such
clusters would be observationally dark matter-free, as their
dynamics would be fully described by baryonic mass alone. This
prediction offers a decisive test to distinguish Tsallis gravity
from both $\Lambda$CDM (which requires dark matter in all
clusters) and MOND (which fails at cluster scales without dark
matter). The observation of even a single dark matter-free galaxy
cluster would be highly unexpected within the standard
cosmological model and would constitute a major paradigm shift in
our understanding of cosmic structure formation.
\begin{table}[htbp]
\centering \caption{Bounds on the Tsallis parameter $\delta$ from
different physical frameworks (see \cite{Luciano2022} and
references therein).}
\begin{tabular}{lll}
\hline
Bound & Physical framework & Reference \\
\hline
$\delta \gtrsim 0.5$ & Galactic rotation curves & \cite{Sheykhi2020} \\
$\delta > 0.5$ & Late-time accelerated expansion & \cite{Sheykhi2020} \\
Running & Boson mixing & \cite{Luciano2021b} \\
$0.993 \lesssim \delta \lesssim 0.994$ & $^7$Li abundance & \cite{Ghoshal2021} \\
$\delta = 1.222$ & Cosmic ray observations & \cite{Beck2009} \\
$0.697 \leq \delta \leq 4.2$ & Entanglement measurements & \cite{Moslehi2020} \\
$|\delta-1| \lesssim 10^{-13}$ & Non-commutative geometry & \cite{Moslehi2020} \\
$\delta \geq 1.203$ & Quark coalescence & \cite{Biro2013} \\
$1 \leq \delta < 1.333$ & High-energy collisions & \cite{Bhattacharyya2017} \\
$\delta \geq 1.218$ & Black holes & \cite{Mejrhit2020} \\
Running & Fermion mixing & \cite{Luciano2021c} \\
$\delta \lesssim 1.222$ & Unruh effect & \cite{Luciano2021d} \\
$\delta \approx 1.01$ & Big Bang Nucleosynthesis & \cite{Ghoshal2021} \\
$0.996 \lesssim \delta \lesssim 1.001$ & $^4$He, $^2$H abundance & \cite{Ghoshal2021} \\
$\delta \gtrsim 0.999$ & Dark Matter relic density & \cite{Ghoshal2021} \\
\hline
\end{tabular}
\label{tab:delta_bounds}
\end{table}
\section{Closing remarks \label{Closing}}
\noindent

The program of entropic gravity, initiated by Jacobson and
Verlinde, aims to derive gravitational dynamics from thermodynamic
principles. A central question has remained: which entropy
functional correctly captures the microscopic degrees of freedom
of spacetime? To answer this question, in this paper, we have
tested all major generalized entropies against observational data-
from galactic rotation curves to X-ray cluster masses. We
disclosed that only Tsallis entropy passes all tests. Other
modified entropies such as R\'{e}nyi, Kaniadakis, Logarithmic, and
Barrow entropies either fail to reproduce flat rotation curves or
predict inconsistent cluster masses. The Tsallis framework, in
contrast, not only recovers Newtonian gravity in the appropriate
limit but also naturally gives rise to MOND through the
zero-energy bits mechanism of Abreu and Neto, and simultaneously
explains the missing mass in galaxy clusters \cite{Abreu2013}.
Thus, the entropic force paradigm, when confronted with empirical
evidence, uniquely selects Tsallis entropy as the correct
statistical foundation for emergent gravity. If gravity is
accepted as an emergent phenomenon rooted in thermodynamic laws,
then Tsallis entropy uniquely provides a formulation that
simultaneously reproduces Newtonian gravity in the classical
limit, MOND theory on galactic scales, and the dynamics of galaxy
clusters without dark matter. Therefore, for the entropic gravity
paradigm to be consistent with observational data across all
scales, from globular clusters to galaxies to galaxy clusters, it
must inevitably be built upon Tsallis entropy. This brings the
entropic gravity program to its natural conclusion: gravity, at
its core, is consistently described by nonextensive statistical
mechanics.

In this work we have systematically confronted the major
generalized entropy proposals namely, R\'{e}nyi, Kaniadakis,
Logarithmic, Barrow, and Tsallis with observational data spanning
from galactic rotation curves to X-ray galaxy cluster masses and
globular cluster velocity dispersions. Our analysis leads to
several clear conclusions. The main results we found can be
summarized as follows.

(i) Failure of type-II entropies: Entropies that are simple
additive corrections to the Bekenstein-Hawking area law
(R\'{e}nyi, Kaniadakis, Logarithmic) cannot reproduce the flat
rotation curves of galaxies. When applied to galaxy clusters, they
produce mass profiles that are either un-physical or inconsistent
with X-ray observations. The same holds for Barrow entropy
(type-I) when its exponent is restricted to the usual range
$0\leq\Delta\leq1$; it either falls off too rapidly or, if the
formalism is misapplied, yields a scaling that does not match
cluster data.

(ii) Tsallis entropy succeeds across all scales: Only Tsallis
entropy, with its single nonextensive parameter $\delta$, passes
every test.

On galactic scales ($\sim10$ kpc), $\delta>1/2 \ (\delta\approx
0.5)$ explains flat rotation curves and recovers the Tully-Fisher
relation.

On galaxy cluster scales ($\sim100$ kpc), we analyzed 40
high-quality clusters and obtained $\langle \delta \rangle =
0.97547 \pm 0.004.$ The predicted mass profiles from the
Tsallis-modified Newtonian force law, $F\propto GMm/r^{2\delta}$
agree remarkably well with the observed X-ray mass distributions.

On globular cluster scales ($\sim0.01$ kpc), using updated
kinematic data for 33 clusters \cite{Baumgardt2018}, we find
$\langle \delta \rangle= 1.0046 \pm 0.0025$. This value is
consistent with Newtonian gravity ($\delta=1$) and explains why
most globular clusters do not require dark matter, while still
accommodating the mild flattening seen in a few clusters (e.g.,
NGC 7078 shows a slight deviation).

\begin{figure}[h]%
\centering
\includegraphics[width=1\textwidth]{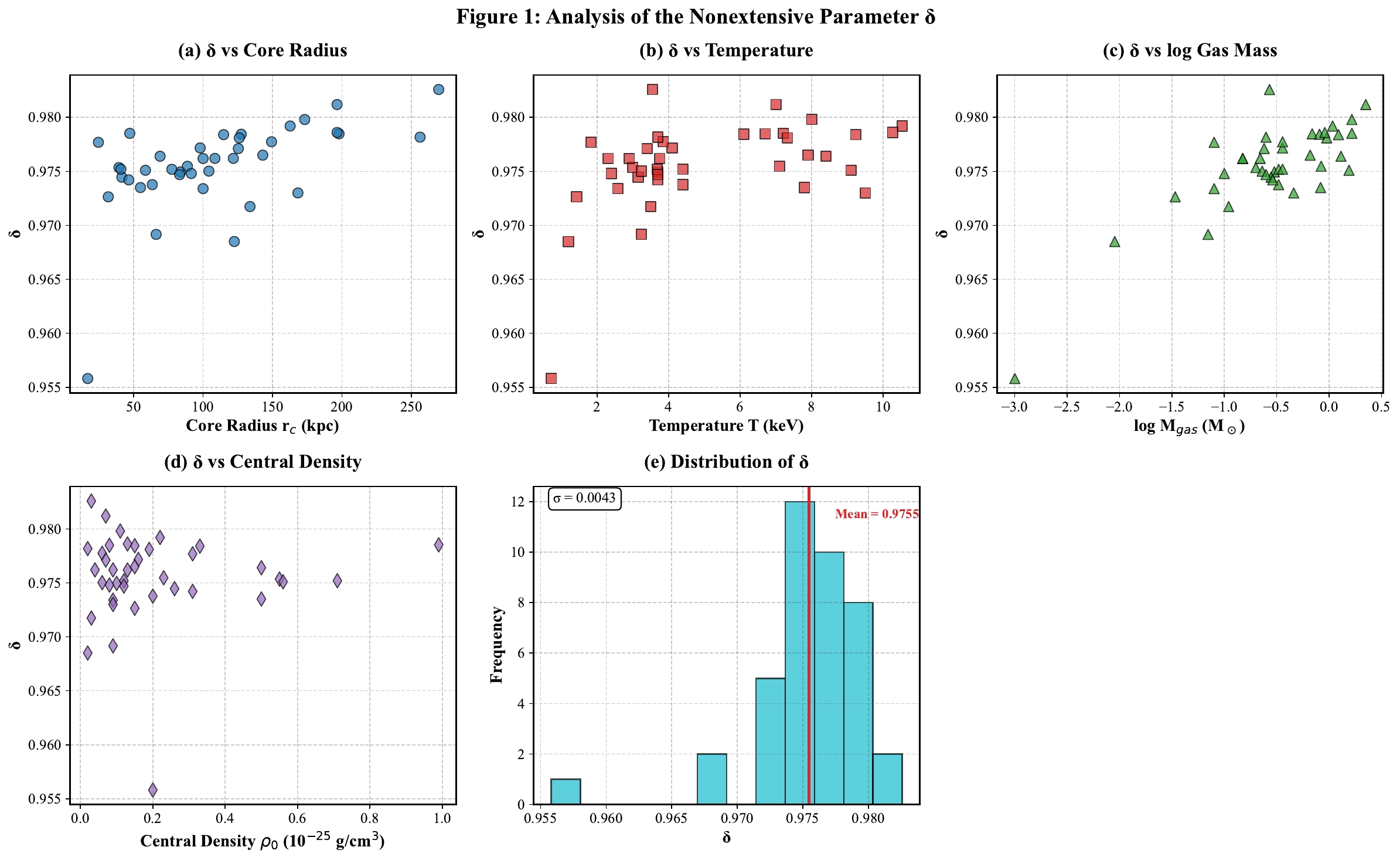}
\caption{ Distribution of the Tsallis nonextensive parameter
$\delta$ for 40 galaxy clusters as a function of mass ($M$) and
temperature ($T$). The scattered points indicate no significant
correlation between $\delta$ and the macroscopic properties of the
clusters.} \label{fig1}
\end{figure}

\begin{figure}[h]%
\centering
\includegraphics[width=0.9\textwidth]{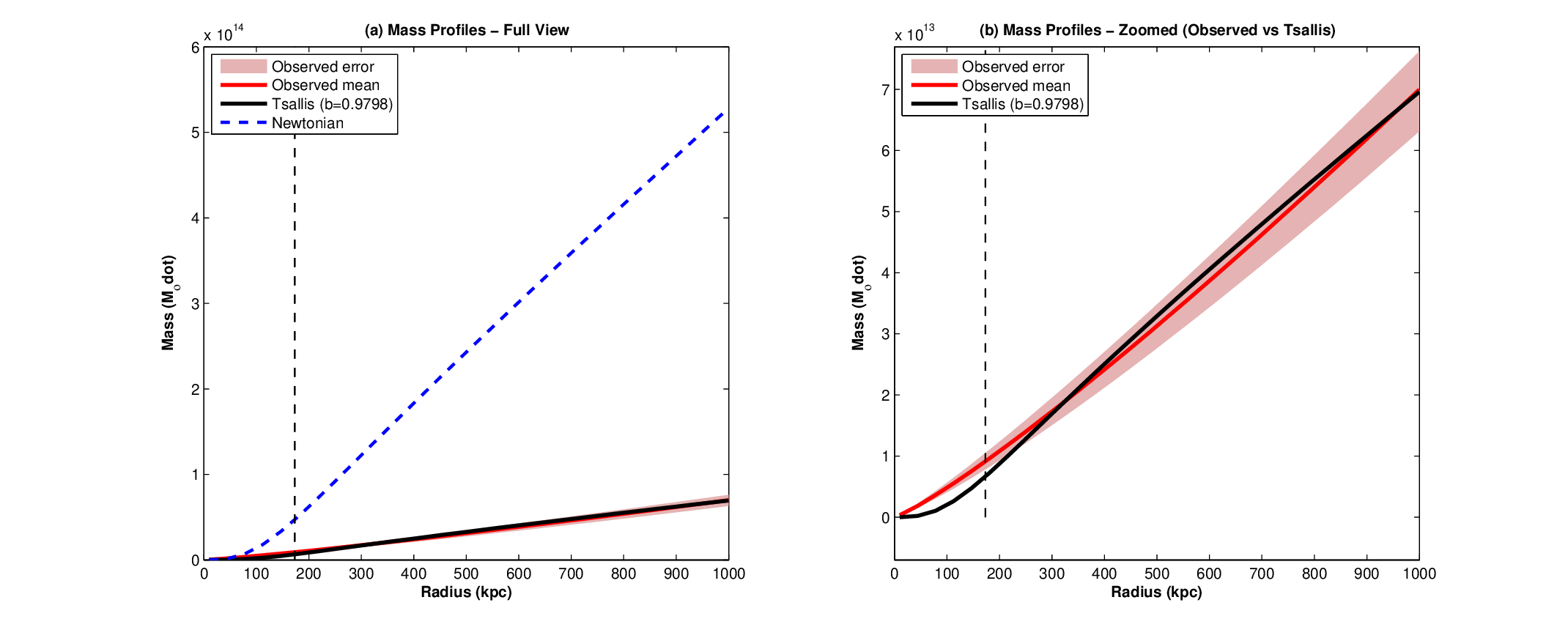}
\caption{The radial mass profile of the galaxy cluster A0401 .
The cluster parameters are $T = 8\times1160415$~K,
$\beta = 0.613$, and $r_c = 173.2\times3.86\times10^{19}$~m,
adopted from Ref.~\cite{Reiprich2002}, with a nonextensive
parameter $\delta = 0.9798$. Panel (a) displays the full view of
the mass profile, including the observational data with its error
band (red shaded region), the observational mean (red solid line),
the Tsallis prediction (black solid line), and the Newtonian
prediction (blue dashed line). In panel (b), we show the same
profile with the Newtonian curve omitted, so that the Tsallis
prediction can be compared directly with the observational data.
The vertical dashed line indicates the core radius $r_c$. } \label{fig5}
\end{figure}

\begin{figure}[h]%
\centering
\includegraphics[width=0.9\textwidth]{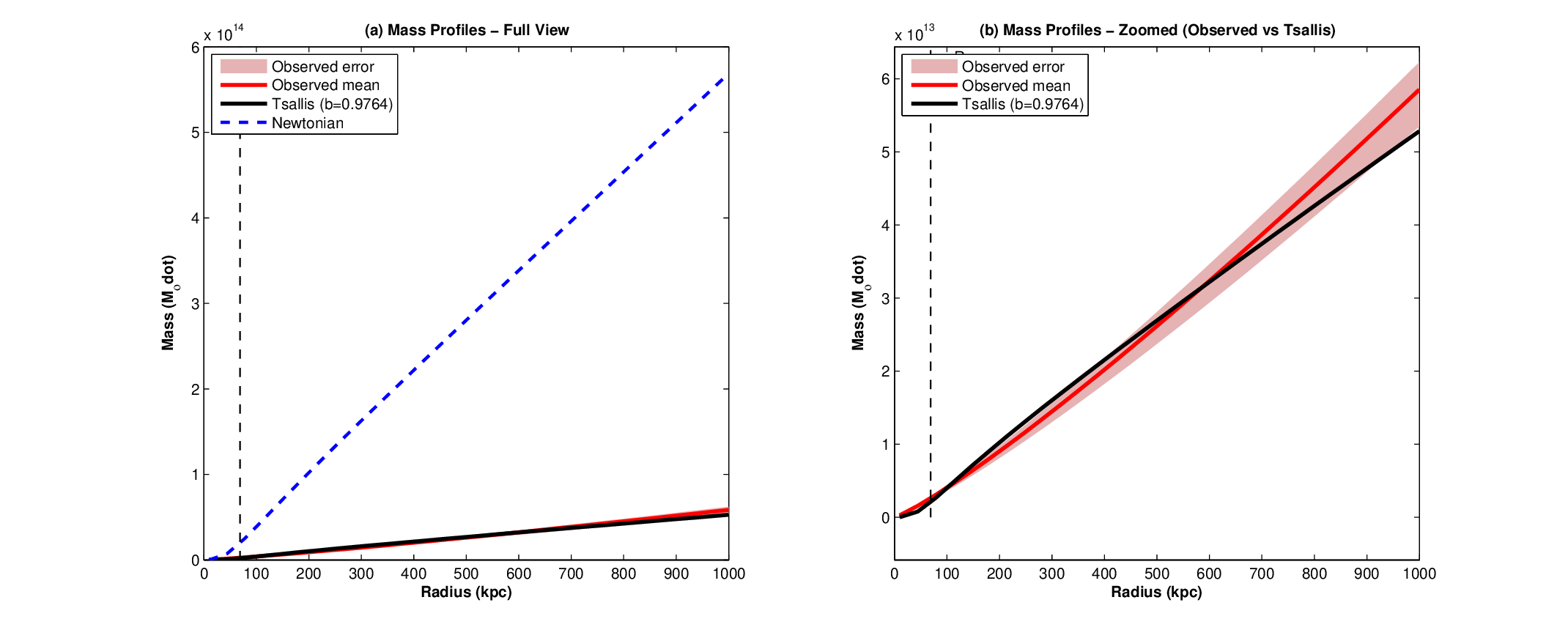}
\caption{The radial mass profile of the galaxy cluster A0478 .
The cluster parameters are $T = 8.4\times1160415$~K,
$\beta = 0.613$, and $r_c = 69\times3.86\times10^{19}$~m,
adopted from Ref.~\cite{Reiprich2002}, with a nonextensive
parameter $\delta = 0.9764$. Panel (a) displays the full view of
the mass profile, including the observational data with its error
band (red shaded region), the observational mean (red solid line),
the Tsallis prediction (black solid line), and the Newtonian
prediction (blue dashed line). In panel (b), the Newtonian curve
is omitted so that the Tsallis prediction can be compared
directly with the observational data. The vertical dashed line
indicates the core radius $r_c$.} \label{fig6}
\end{figure}

(iii) Scale dependence of $\delta$ without within-scale
correlation: The Tsallis parameter takes distinct values on
different scales: $\delta>1/2 \ (\delta\approx 0.5)$ (galaxies),
$\delta \approx 0.97547$ (clusters), $\delta \approx 1.005$
(globular clusters). However, within each scale, $\delta$ shows no
correlation with mass, radius, temperature, or any other
instantaneous system property (see Fig.~\ref{fig1}).
This indicates that $\delta$ is not determined by the present-day
macroscopic state of the system, but rather encodes the cumulative
effects of its dynamical history, including age, relaxation state,
and environmental interactions, which manifest as different
effective non-extensivities at each characteristic scale. This
interpretation is fully consistent with the findings of Carvalho
et al. \cite{Carvalho2007} and Komatsu \cite{Komatsu2012}, who
demonstrated that the Tsallis parameter correlates with cluster
age and dynamical relaxation state.

(iv) Prediction of dark-matter-free galaxy clusters: Because
$\delta$ can be as low as 0.973 in clusters and as high as 1.005
in globular clusters, the continuity of the framework implies that
clusters with $\delta=1$ (i.e., Newtonian behaviour) should exist.
Such clusters would have dynamical masses exactly equal to their
baryonic mass, containing no dark matter. This is a sharp,
testable prediction that distinguishes Tsallis gravity from both
$\Lambda$CDM (which requires dark matter in all clusters) and MOND
(which fails at cluster scales without additional dark matter).
The discovery of even one such cluster would provide decisive
evidence for the Tsallis paradigm.

(v) Theoretical consistency: Tsallis entropy arises naturally from
nonextensive statistical mechanics, which is required for systems
with long-range interactions (like gravity) and for black hole
thermodynamics where the area law implies sub-extensivity. Unlike
the von Neumann entropy, Tsallis entropy provides a well-defined
thermodynamic entropy in the emergent gravity framework, resolving
the Carroll-Remmen criticism that standard entanglement entropy is
not a valid basis for entropic gravity. Moreover, Tsallis gravity
can reproduce MOND in the appropriate limit via the zero-energy
bits mechanism, while simultaneously explaining cluster masses-
something MOND alone cannot do.

(vi) Uniqueness of Tsallis entropy: Given that all other known
generalized entropies fail at least one of the observational tests
(galactic rotation curves or cluster masses), we conclude that
Tsallis entropy is the unique entropic foundation for a consistent
emergent gravity theory that describes gravitational phenomena
from kiloparsec to megaparsec scales without invoking dark matter.

Many issues remain for future investigations. One can extend the
analysis to larger samples of clusters and globular clusters with
improved kinematic data. Besides, one may derive the full
cosmological implications of the scale-dependent $\delta$ and
compare with CMB and large-scale structure observations.
Furthermore, one may try to understand the microscopic origin of
the discrete $\delta$ values-perhaps as fixed points of a
renormalisation group flow for gravity. Finally, the predicted
$\delta=1$ clusters should be searched for in existing X-ray and
weak-lensing surveys: their existence would revolutionise our
understanding of cosmic structure.

In summary, the entropic gravity programme, when confronted with
multi-scale astrophysical data, uniquely selects Tsallis
nonextensive statistics as the correct thermodynamic description
of spacetime. Gravity, at its most fundamental level, appears to
be an emergent phenomenon rooted in nonextensive statistical
mechanics.
\acknowledgments{We thank Shiraz University Research Council.}

\newpage

\begin{figure}[h]%
\centering
\includegraphics[width=0.9\textwidth]{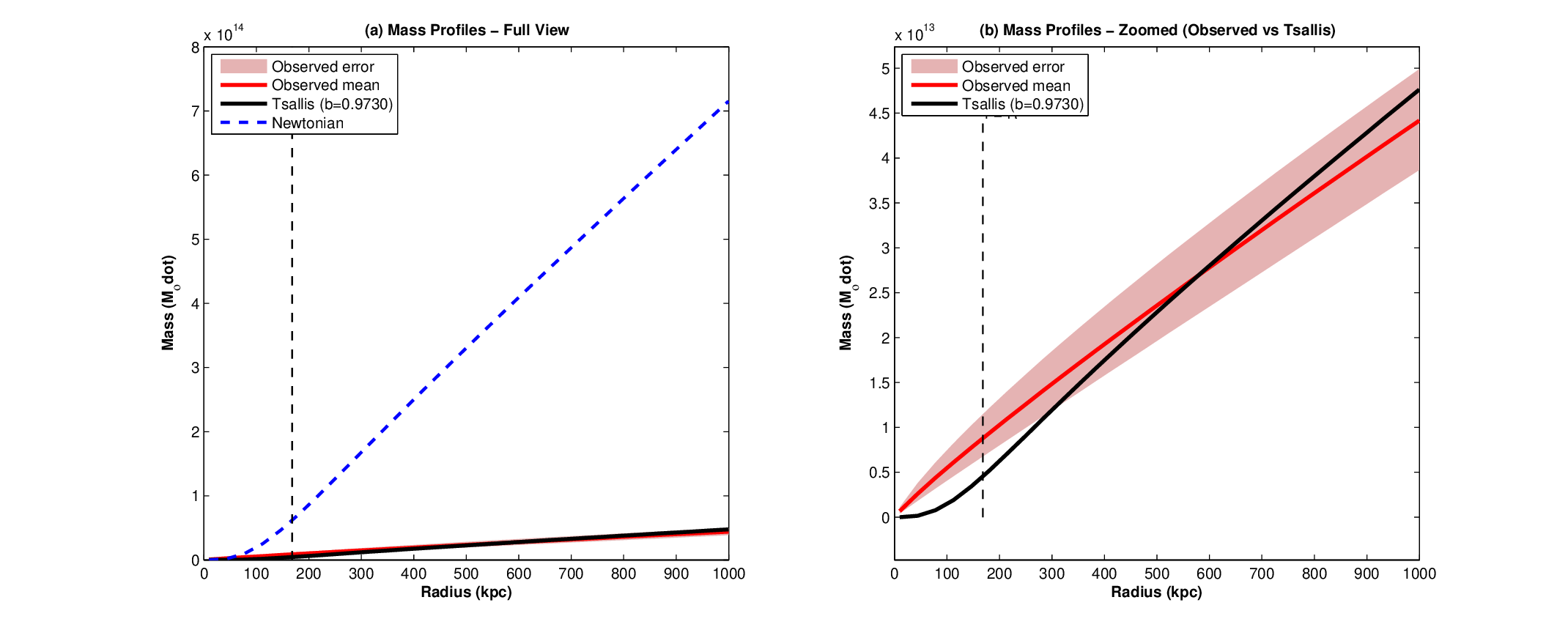}
\caption{ The radial mass profile of the galaxy cluster A0754 .
 The cluster parameters are $T = 9.5\times11604525$~K,
$\beta = 0.698$, and $r_c = 168.3\times3.86\times10^{19}$~m,
adopted from Ref.~\cite{Reiprich2002}, with a nonextensive
parameter $\delta = 0.9730$. Panel (a) displays the full view of
the mass profile, including the observational data with its error
band (red shaded region), the observational mean (red solid line),
the Tsallis prediction (black solid line), and the Newtonian
prediction (blue dashed line). In panel (b), the Newtonian curve
is omitted so that the Tsallis prediction can be compared
directly with the observational data. The vertical dashed line
indicates the core radius $r_c$. } \label{fig7}
\end{figure}

\begin{figure}[h]%
\centering
\includegraphics[width=0.9\textwidth]{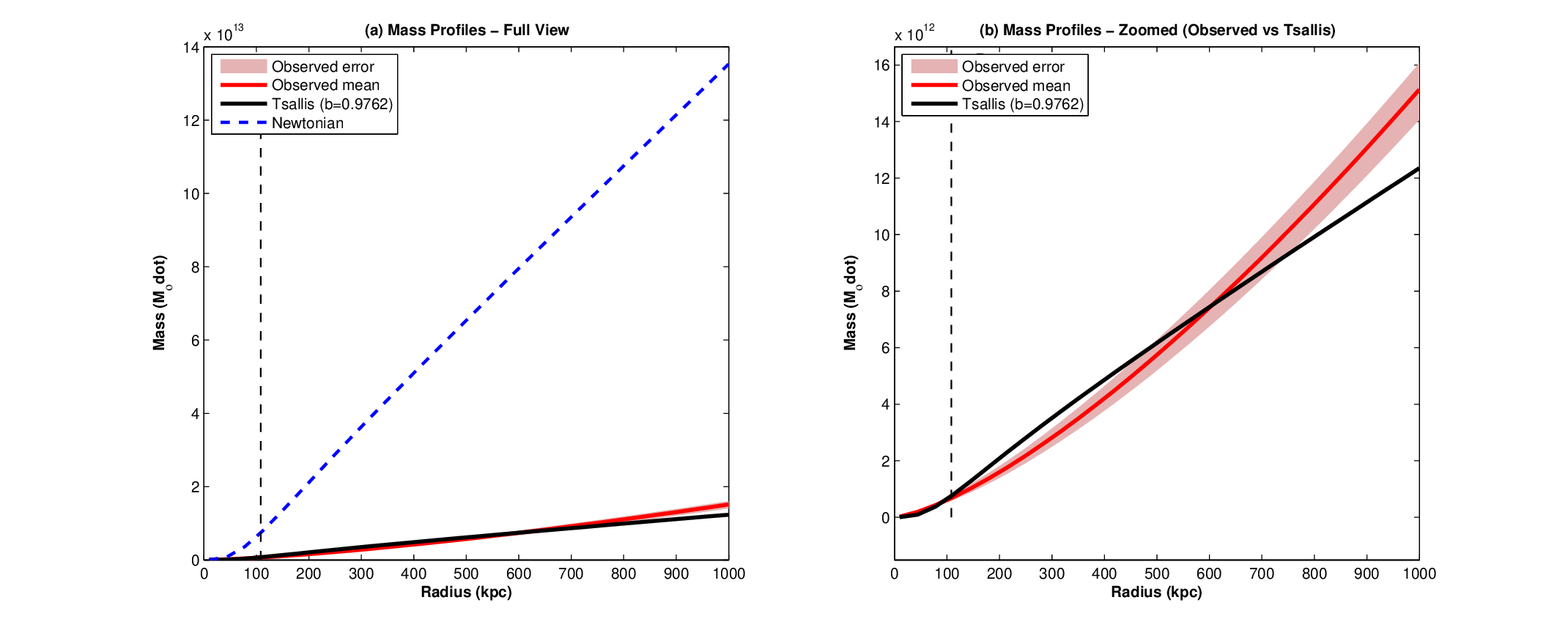}
\caption{The radial mass profile of the galaxy cluster A0400
. The cluster parameters are $T = 9.7\times1160415$~K,
$\beta = 0.591$, and $r_c = 108.5\times3.86\times10^{19}$~m,
adopted from Ref.~\cite{Reiprich2002}, with a nonextensive
parameter $\delta = 0.9762$. Panel (a) displays the full view of
the mass profile, including the observational data with its error
band (red shaded region), the observational mean (red solid line),
the Tsallis prediction (black solid line), and the Newtonian
prediction (blue dashed line). In panel (b), the Newtonian curve
is omitted so that the Tsallis prediction can be compared
directly with the observational data. The vertical dashed line
indicates the core radius $r_c$. } \label{fig8}
\end{figure}

\begin{figure}[h]%
\centering
\includegraphics[width=0.6\textwidth]{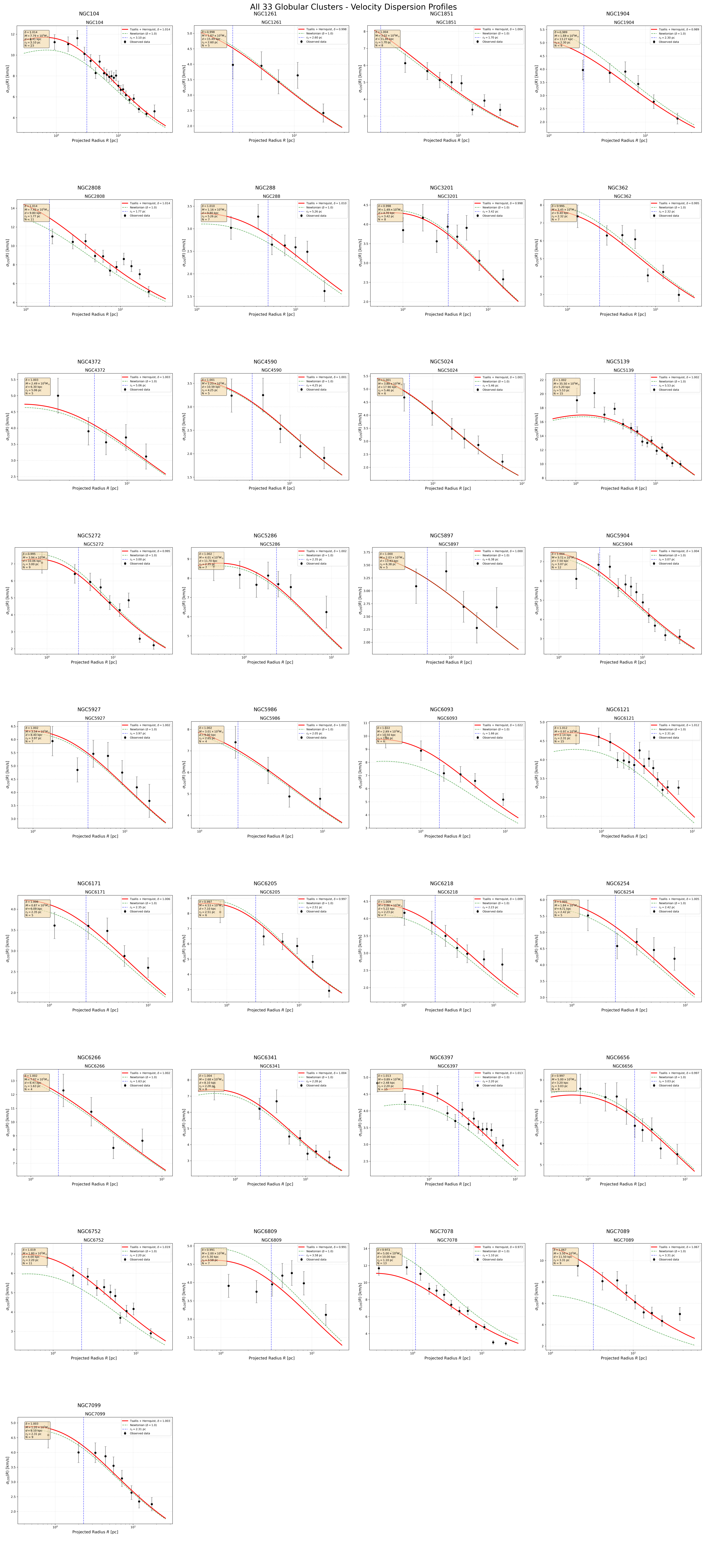}
\caption{Line of sight velocity dispersion profiles for selected globular clusters fitted with the Tsallis modified gravity model. The solid curves represent the best-fit Tsallis predictions with the corresponding $\delta$ values (see Table II), while the data points are from the Baumgardt \& Hilker (2018) catalog. }
\label{fig9}
\end{figure}
\end{document}